\documentstyle[12pt,epsfig,epsf]{article}
\def\ga{\mathrel{\raise.3ex\hbox{$>$\kern-.75em\lower1ex\hbox{$\sim$}}}}
\def\la{\mathrel{\raise.3ex\hbox{$<$\kern-.75em\lower1ex\hbox{$\sim$}}}}

\begin{document}

\vspace*{0.6cm}
\begin{center}
{\large{{\bf Flavor Changing Neutral Currents involving Heavy Quarks\\
 with Four Generations}}}

\vspace{1.cm}
Abdesslam Arhrib$^{1,2}$ and Wei-Shu Hou$^{3}$ \\
$^{1}$National Center for Theoretical Sciences,
PO-Box 2-131 Hsinchu, Taiwan 300\\
$^{2}$Facult\'e des Sciences et Techniques B.P 416 Tangier, Morocco\\
$^{3}$Department of Physics, National Taiwan University, Taipei,
Taiwan 106

\end{center}

\vspace{1.cm}
\begin{abstract}
We study various flavor changing neutral currents (FCNC) involving
heavy quarks in the Standard Model (SM) with a sequential fourth
generation. After imposing $B\to X_s\gamma$, $B\to X_sl^+l^-$ and
$Z\to b\bar{b}$ constraints, we find ${\cal B}(Z\to
s\bar{b}+\bar{s}b)$ can be enhanced by an order of magnitude to
$10^{-7}$, while $t\to cZ, cH$ decays can reach $10^{-6}$, which
are orders of magnitude higher than three generation SM. However,
these rates are still not observable for the near future.
With the era of Large Hadron Collider approaching, we focus on
FCNC decays involving fourth generation $b^\prime$ and $t^\prime$
quarks. We calculate the rates for loop induced FCNC decays
$b^\prime\to bZ,\ bH,\ bg,\ b\gamma$, as well as $t^\prime\to tZ,\
tH,\ tg,\ t\gamma$. If $|V_{cb'}|$ is of order $|V_{cb}| \simeq
0.04$, tree level $b^\prime\to cW$ decay would dominate, posing a
challenge since $b$-tagging is less effective. For $|V_{cb'}| \ll
|V_{cb}|$, $b'\to tW$ would tend to dominate, while $b'\to
t^\prime W^*$ could also open for heavier $b'$, leading to the
possibility of quadruple-$W$ signals via $b'\bar b'\to b\bar b
W^+W^-W^+W^-$. The FCNC $b'\to bZ, bH$ decays could still dominate
if $m_{b'}$ is just above 200 GeV. For the case of $t'$, in
general $t^\prime\to bW$ would be dominant, hence it behaves like
a heavy top. For both $b'$ and $t'$, except for the intriguing
light $b'$ case, FCNC decays are typically in the $10^{-4} -
10^{-2}$ range, and are quite detectable at the LHC.
For a possible future International Linear Collider, we find the
associated production of FCNC $e^+e^-\to b\bar s$, $t\bar c$ are
below sensitivity, while $e^+e^-\to b^\prime\bar b$ and
$t^\prime\bar t$ can be better probed.
Tevatron Run-II can still probe the lighter $b'$ or $t'$ scenario.
LHC would either discover the fourth generation and measure the
FCNC rates, or rule out the fourth generation conclusively. If
discovered, the ILC can study the $b'$ or $t'$ decay modes in
detail.
\end{abstract}

\newpage
\section{Introduction}

The successful Standard Model (SM) of electroweak interactions is
a renormalizable theory, but it still could be just an effective
theory of a more fundamental or more complete theory that is yet
to be discovered. The goal of the next generation of high energy
colliders, such as the Large Hadron Collider (LHC) \cite{LHC}, or
the International Linear Collider (ILC) \cite{ILC}, %TESLA,GLC},
is to probe the origins of electroweak symmetry breaking, and/or
discover new physics phenomena.

The apparent suppression of flavor changing neutral currents
(FCNC) has played a critical role in the establishment of the
three generation SM.
With just one Higgs doublet, by unitarity of the quark mixing
matrix, the couplings of neutral bosons (such as the $Z$ boson,
Higgs boson, gluon and photon) to a pair of quarks are flavor
diagonal at the tree level. At the one loop level, the charged
currents (CC) do generate FCNC $Q\to q\{Z, H, g, \gamma\}$
transitions, but they are suppressed by the GIM mechanism.
Interesting phenomena such as $CP$ violation that are the current
focus at the B factories involve FCNC $b\to s$ transitions.
Because of very strong GIM cancellation between $d$, $s$ and $b$
quark loops, and in part because of an unsuppressed decay width,
however, the corresponding top decays are rather suppressed
\cite{sm1,sm2} in the three generation SM, viz.
\begin{eqnarray}
& & {\cal B}(t \to cZ) = 1.3 \times 10^{-13},\
 {\cal B}(t \to cH) = (5.6 - 3.2) \times 10^{-14},
 \label{eq1}\\
& & {\cal B}(t \to cg) =  3.8 \times 10^{-11}, \
 {\cal B}(t \to c \gamma) = 4.3 \times 10^{-13},
 \label{eq2}
\end{eqnarray}
where $M_H =$ (115 -- 130) GeV has been taken. Together with the
expectation ${\cal B}(Z \to b\bar{s}+\bar{b}s) \cong 10^{-8}$,
such strong suppression within three generation SM implies that
these processes are excellent probes for new physics, such as
supersymmetry, extended Higgs sector, or extra fermion families.

In the last decade, there has been intense activities to explore
FCNC involving the top quark. Experimentally, CDF, D0 \cite{cdfd0}
and LEPII \cite{lep2} collaborations have reported interesting
bounds on FCNC top decays. These bounds are rather weak, however,
but will improve in the coming years, first with Tevatron Run II,
in a few years with the LHC, and eventually at the ILC.
The expected sensitivity to top FCNC at Tevatron Run II is about
${\cal B}(t\to c \gamma ) \ga 5 \times 10^{-3}$, while at the LHC,
with one year of running, it is possible to probe the range
\cite{LHC,aguila},
\begin{eqnarray}
& & {\cal B}(t\to c Z ) \ga 7.1\times 10^{-5}, \ \
 {\cal B}(t\to c H) \ga 4.5\times 10^{-5},
 \nonumber\\
& & {\cal B}(t\to cg) \ga 10^{-5}, \ \
 {\cal B}(t\to c \gamma ) \ga 3.7\times 10^{-6}.
 \label{LHC}
\end{eqnarray}
At the ILC, the sensitivity is slightly less \cite{aguila}, and
the range
\begin{eqnarray}
{\cal B}(t\to c H) \ga 4.5\times 10^{-5}, \ \
 {\cal B}(t\to c \gamma) \ga 7.7\times 10^{-6},
 \label{ILC}
\end{eqnarray}
can be probed. Thus, models which can enhance these FCNC rates and
bring them close to the above sensitivities are welcome.

From the theoretical side, many SM extensions predict that top and
$Z$ FCNC rates can be orders of magnitude larger than their SM
values (see Ref. \cite{exof} for a review).
The aim of this paper is to study FCNC involving heavy quarks in
the more modest extension of SM by adding a sequential fourth
generation. This retains all the features of SM, except bringing
into existence the heavy quarks $b'$ and $t'$. We will first cover
the impact on FCNC top decays: $t\to cZ,\ cH,\ cg,\ c\gamma$. We
find that a 4th generation still cannot bring these decay rates to
within experimental sensitivity, once constraints from rare $B$
decays are imposed. Likewise, as discussed in a later section,
FCNC decays of the Z boson, e.g. $Z\to \bar{b}s+\bar{s}b$, also
remain difficult.

With LHC in view, however, it is timely to address the decay and
detection of 4th generation quarks $b^\prime$ and $t^\prime$
themselves. For this matter, we include in our analysis both CC
decays as well as FCNC $b^\prime\to bX$ and $t^\prime\to tX$
decays, and evaluate the total widths and branching ratios. We
illustrate the search strategies at the LHC (and Tevatron Run-II),
and find that FCNC decays are typically at the
$10^{-4}$--$10^{-2}$ order, and should be detectable at the LHC.
For a relatively light $b'$ just above 200 GeV, FCNC $b'$ decays
could still dominate! For completeness, we also study the
signature of these FCNC couplings at $e^+e^-$ colliders through
heavy and light quark associated production, i.e. FCNC $e^+e^-\to
\bar bs+\bar sb$, $\bar{t}c+\bar{c}t$, $\bar{b}^\prime b
+\bar{b}{b}^\prime$ and $\bar{t}^\prime t+\bar{t}{t}^\prime$.

Before we turn to our detailed study, we should address the issue
as to whether a 4th generation itself is relevant at all. Let us
first start with the original motivation for considering a 4th
generation.
Despite its great success, the 3 generation Standard Model (SM3)
may be incomplete. The generation structure (including the
question of ``Why 3?") is simply not understood, while having just
one heavy quark, the top, with mass at the weak scale is also
puzzling. A simple enlargement of SM3 by adding a sequential
fourth generation (SM4), where we have already used the notation
of $t^\prime , b^\prime$ for the fourth generation quarks, could
allow us to gain further insight into the question of flavor.

The question that immediately arises is neutrino counting via the
invisible $Z$ width~\cite{PDG}. Indeed, one of the original strong
motivations for the 4th generation, before the advent of the
$N_\nu$ result from SLD and LEP experiments, was the possibility
of the 4th neutral lepton as a dark matter (DM) candidate. But
with active neutrino number $N_\nu$ convincingly established at 3,
since 1989 the 4th generation as a whole fell out of favor. Recent
observations of neutrino oscillations, however, point toward an
enlarged neutrino sector~\cite{sher,evseesaw1,evseesaw2}, which by
itself must be beyond minimal SM3. For example, it has been
demonstrated recently~\cite{evseesaw3} that, if the right-handed
neutrino Majorana mass scale is of the order ${\cal O}$ (eV),
dubbed ``eV seesaw", a fit to LSND data \cite{LSND} can be
obtained. Within such eV seesaw scenario, unlike the very high
scale standard seesaw, extension to the fourth generation can be
easily accomplished. In a recent work \cite{evseesaw1}, it has
been shown that such eV seesaw can be extended to four lepton
generations. The 4th ``neutrino" or neutral lepton $N$ is
pseudo-Dirac and heavy, so it does not affect the invisible $Z$
width. The 3 extra sterile neutrinos allow one to accommodate LSND
data. Taking this as a plausible scenario that $N_\nu = 3$ is no
longer an impediment to having a 4th generation, we will not
discuss the lepton sector any further in this work.

The second, rather serious issue to face is about precision
electroweak constraints, which seem to pose a challenge to the
fourth generation. In particular, ``an extra generation of
ordinary fermions is excluded at the 99.95\% CL on the basis of
the $S$ parameter alone"~\cite{PDG}, and efforts to reduce $S$
tend to increase $T$. With $S = -0.04 (-0.09)\pm 0.11$, $T = -0.03
(+0.09)\pm 0.13$ for $m_H = 100 (300)$ GeV, the problem is serious
for many extensions beyond SM3, such as technicolor, or
SM4~\cite{EW2005}. Even so, for SM4, if the extra neutrino is
close to its direct mass limit of $m_N \ga M_Z/2$, this can drive
$S$ smaller at the expense of a larger $T$, and a more detailed
analysis suggest a 4th generation is not ruled out~\cite{he}.

If a 4th generation exists, it will appear soon. The effort of
this paper is to update $b'$ and $t'$ decays, including the often
dominant charged current decay, to facilitate the search program
at Tevatron and LHC. The LHC would either discover the fourth
generation and measure some the FCNC rates, or rule out the fourth
generation conclusively.
In this vein, we mention that there are possible hints for the 4th
generation from FCNC and $CP$ violating (CPV) $b\to s$ transitions
at the B factories. The difference in measured~\cite{AKpi0} direct
CPV in $B\to K^+\pi^-$ and $K^+\pi^0$ modes could arise from New
Physics phase in the electroweak penguin process, and the 4th
generation is an excellent candidate~\cite{HNS}. Furthermore, the
well-known hint~\cite{HFAG} of a difference between mixing
dependent CPV measurements in a host of charmless $b\to s\bar qq$
type of modes vs $B \to J/\psi K_S$, dubbed the $\Delta S$
problem, could also be partially explained by 4th generation CPV
effect through electroweak penguin amplitude~\cite{HNRS}. We
conclude that the fourth generation is not ruled out, may be
hinted at in FCNC/CPV $b\to s$ transition data, and $b'$ and $t'$
search at the Tevatron and especially the LHC is imperative.

The paper is organized as follows. In the next section we  review
the experimental constraints from electroweak precision tests,
$Z\to \bar bb$, and from FCNC $b\to s$, $s\to d$ and $c\to u$
transitions. In Section 3 we study FCNC $t\to c X$ decays, and in
Section 4 $b^\prime \to bX$, $t^\prime \to tX$ decays, where $X=Z,
H, g$ and $\gamma$. In Section 5 we investigate the associated
production $e^+e^-\to \bar{q}Q+\bar{Q}q$, for $(Q, q) = (b, s)$,
$(t, c)$, $(b^\prime, b)$ and $(t^\prime, t)$ at (Super)B
Factories, possible GigaZ, and the future ILC collider. After some
discussions in Section 6, our conclusions are given in Section 7.
An Appendix deals briefly with suppressed FCNC $b'\to sX$ and
$t'\to cX$ decays.

%%%%%%%%%%%%%%%%%%%%%%%%%%%%%%%%%%%%%%%%%%%%%%%%%%%%%%%%%%%%%%%%
\section{Constraints}

With additional quark mixing elements, one crucial aspect is that
the source for $CP$ violation (CPV) is no longer unique.
Consideration of CPV phases is important, as it enlarges the
allowed parameter space from low energy considerations~\cite{AH2}.
Existing experimental data as well as theoretical arguments put
stringent constraints on the masses and mixings involving the
fourth generation, some of which will be reviewed here.

\subsection{CKM Unitarity}

Adding a fourth family enlarges the CKM quark mixing matrix, and
the present constraint on the various CKM elements $V_{ij}$ for
SM3 (i.e. $i,\ j = 1$--$3$) that are known only indirectly, are
considerably relaxed \cite{PDG}. Put in other words, flavor
physics data do not preclude a 4th generation. For example, the
elements $|V_{ts}|$ and $|V_{td}|$ can be as large as about 0.11
and 0.08, respectively \cite{PDG}.
Constraints on CKM elements involving the 4th generation is rather
weak. For example, unitarity of the first row of $V$ allows
$|V_{ub^\prime}| < 0.08$~\cite{PDG,Vubp}. In fact, the long
standing puzzle of (some deficit in) unitarity of the first row
could be taken as a hint for finite $\vert V_{ub^\prime}\vert$.

\subsection{Direct Search}

Experimental search for fourth generation quarks has been
conducted by several experiments. These experimental searches
clearly depend strongly on the decay pattern of the fourth
generation. A strict bound on $b^\prime$ mass comes from LEP
experiments, $m_{b'}\ga M_Z/2$ GeV \cite{LEP}, where both CC and
FCNC decays of $b^\prime$ has been considered.

At the Tevatron, where the heavy top quark was discovered, both CDF and
D0 have searched for fourth generation quarks.
The top quark search applies to $b'$ and $t'$ quarks that decay
predominantly into $W$ (i.e. $b'\to c W$ and $t'\to b W$), and the
corresponding lower bounds are $m_{t',b'}\ga 128$ GeV~\cite{D0}.
Searching for $b^\prime$ quark through its FCNC decays, an
analysis by D0 excludes the $b'$ in the mass range $M_Z/2 < m_{b'}
<M_Z+m_b$ that decays via the FCNC $b'\to b \gamma$
process~\cite{D01}. CDF excludes \cite{cdf4} the $b'$ quark in the
mass range of 100 GeV $< m_{b'}< 199 $ GeV at 95\% C.L., if ${\cal
B}(b'\to b Z)=100\%$. Assuming ${\cal B}(b'\to b Z)=100\%$, CDF
also excluded long-lived $b'$ quark with mass in the range
$M_Z+m_b < m_{b^\prime} < 148$ GeV and a lifetime of $3.3\times
10^{-11}$ s \cite{cdf148}.
Recently, CDF has looked for long-lived fourth generation quarks
in a data sample of 90 pb$^{-1}$ in $\sqrt{s}=1.8$ TeV $\rm
p\bar{\rm p}$ collisions, by using signatures of muon-like
penetration and anomalously high ionization energy loss. The
corresponding lower bounds are $m_{t'}\ga 220$ and $m_{b'}\ga 190$
GeV~\cite{acosta}.

The above limits can be relaxed if we consider the possibility
that $b'\to b H$ , $b'\to c W$ and $b'\to b Z$ decays can be of
comparable size under certain conditions of the CKM
elements~\cite{AH,santos}. Unless associated CKM elements are
extremely small, in general the $b^\prime$ (and certainly the
$t^\prime$) quark should not be very long-lived.

\subsection{Electroweak and \boldmath $Z\to b\bar b$ Constraints}

Theoretical considerations of unitarity and vacuum stability can
put limits on the masses of the fourth generation
quarks~\cite{sher,unit}. Assuming that the fourth family is close
to degenerate, perturbativity requires $m_{t',b'}\la 550$ GeV.

Even so, for $b^\prime$ and $t^\prime$ below several hundred GeV
that we consider, the $\rho$ parameter provides a significant
constraint on the splitting between $t'$ and $b'$,
$|m_{t'}-m_{b'}|\la M_W$ \cite{sher,PDG,AH}.
We note that, having the extra neutral lepton close to the current
bound of $M_Z/2$ could be accommodated by electroweak
data~\cite{he}, which has now moved~\cite{ADLO2005} much more
favorably towards a 4th generation, as discussed in the
Introduction.

The $Z$ width is now well measured and provides a good constraint.
In particular, $t^\prime$ can contribute to $Z\to b\bar{b}$ at one
loop level. For fixed $m_{t^\prime}$ mass, $V_{t^\prime b}$ can be
constrained. Following Ref.~\cite{yanir,HNS,HNS_KL}, for $m_{t'}=300$ GeV
we have,
\begin{eqnarray}
|V_{tb}|^2 + 3.4 |V_{t^\prime b}|^2  \leq 1.14, \label{Zbb}
\end{eqnarray}
which leads to the bound $|V_{t^\prime b}| \la 0.2$ if we assume
that $V_{tb}\approx 1$. The bound is nontrivial, reducing the
range for FCNC $b'\to b$ and $t'\to t$ rates. But the bound can be
relaxed if $t^\prime$ is close to the top quark in mass. In such
case, we may still expect large mixing between the third and
fourth generation $\sin\theta_{34}\approx {\cal O}(1)$.

\subsection{ \boldmath $B\to X_s\gamma $}

Let us now consider the constraints from $b\to s$ transitions such
as $B\to X_s\gamma$ \cite{HSS} and $B\to X_sl^+l^-$ \cite{HWS}. To
study these transitions, we will not use any particular
parameterization of the CKM matrix. We will instead argue how
large could be the CKM matrix elements that contribute to $b\to s$
transitions. The unitarity of the $4\times 4$ CKM matrix leads to
$\lambda_u + \lambda_c + \lambda_t + \lambda_{t'}=0$, where
$\lambda_f= V_{fs}^* V_{fb}$.
In Ref.~\cite{AH2}, the effect of a sequential fourth generation
on $b\to s$ transitions has been studied taking into account the
presence of a new $CP$ phase in $\lambda_{t'}$. Since $\lambda_u =
V_{us}^*V_{ub}$ is very small in strength compared to the others,
while $\lambda_c = V_{cs}^*V_{cb}\approx 0.04$, we parameterize
$\lambda_{t^\prime} \equiv r_{sb} \, e^{i \phi_{sb}}$~\cite{HNS},
where $\phi_{sb}$ is a new $CP$ phase. Then
\begin{eqnarray}
\lambda_t\cong -0.04 - r_{sb} \, e^{i \phi_{sb}}. \label{ltp}
\end{eqnarray}
With more than three generations and at 90\% C.L. \cite{PDG}, the
range for $\lambda_t=V_{ts}^*V_{tb}$ is from 0 to 0.12. Such
constraint together with Eq.~(\ref{ltp}) give a bound on $r_{sb}$
which is $r_{sb} \la 0.08$ for all CPV phase $\phi_{sb} \in [0,2
\pi]$. As we shall soon see, such large value for $r_{sb}$ may be
in conflict with ${\cal B}(B\to X_s l^+l^-)$.

Using Eq.~(\ref{ltp}) and taking into account the GIM mechanism,
the amplitude for $b\to s$ transitions such as $b\to s \gamma$ and
$Z\to s\bar{b}$ can be written as,
\begin{eqnarray}
{\cal M}_{b\to s}\propto 0.04
[f(m_c)-f(m_t)]+r_{sb}e^{i\phi_{sb}}[f(m_{t^\prime})-f(m_t)],
\label{asb}
\end{eqnarray}
where $f$ is a shorthand for a complicated combination of loop
integrals. The relative sign represents the GIM cancellation
between top and charm, and $t^\prime$ and top. The usual top
effect is located in the first term on the r.h.s. of
Eq.~(\ref{asb}), while the genuine $t^\prime$ effect is contained
in the second term. It is clear that if $\phi_{sb}=0$, $B\to
X_s\gamma$ will be saturated more quickly. However,  for
nonvanishing and large CPV phase $\phi_{sb}$, the $t$ and
$t^\prime$ effects could add in quadrature and the bound becomes
more accommodating. Furthermore, the heavy quark mass dependence
in the loop integral $f(m_Q)$, $Q = t, t^\prime$, is mild, so GIM
cancellation in $f(m_{t^\prime})-f(m_t)$ is rather strong, hence
again accommodating.

The present world average for inclusive $b\to s \gamma$ rate is
\cite{PDG} ${\cal B}(B\to X_s \gamma) = (3.3\pm 0.4)\times
10^{-4}$. Keeping the $B\to X_s\gamma$ branching ratio in the
2$\sigma$ range of (2.5--4.1) $\times 10^{-4}$ in the presence of
the fourth generation, with $m_{t^\prime}=300$ GeV ($400$ GeV), we
have checked that the allowed range for $r_{sb}$ for all CPV phase
$\phi_{sb}$ is $r_{sb} \in [0,0.09]$ ($[0,0.06]$). The allowed
range for $r_{sb}$ reduces to 0.03 (0.02) if we allow only
1$\sigma$ deviation for $b\to s\gamma$. But for $\phi \sim \pi/2$,
$3\pi/2$, when $V_{t's}^*V_{t'b}$ is largely imaginary, $r_{sb}$
can take on much larger values~\cite{AH2}. Thus, the $B\to
X_s\gamma$ constraint is not much more stringent than the general
CKM constraint.

\subsection{\boldmath $B\to X_sl^+l^- $}

The inclusive semileptonic $B\to X_s l^+ l^-$ decay \cite{AH2} has
been measured recently by both Belle and BaBar \cite{Belle,BaBar},
with rates at $(6.1_{-1.8}^{+2.0}) \times 10^{-6}$ and $(5.6 \pm
1.5 \pm 0.6 \pm 1.1) \times 10^{-6}$ respectively. Both Belle and
BaBar have used cuts on the dilepton mass to reduce $J/\psi$ and
$\psi^\prime$ charmonium background. For the muon mode, the cuts
are $-0.25\ {\rm GeV } < M_{\mu\mu}    -M_{J/\psi} < 0.10\ {\rm
GeV }$ and $-0.15\ {\rm GeV}  < M_{\mu\mu} -M_{\psi(2S)} < 0.10\
{\rm GeV} $. For the electron mode, the cut is $m_{e^+e^-}>0.2$
GeV.

With these cuts, for $m_{t'}=300$ (400) GeV and allowing $B\to X_s
l^+ l^-$ to be within the $2\sigma$ range of the experimental
value of $(6.1_{-1.8}^{+2.0}) \times 10^{-6}$, we find that
$r_{sb} \la 0.02$ ($0.01$) for all CPV phase $\phi_{sb}$. The more
stringent bounds apply, however, for $\phi_{sb}\approx \pi$ when
$t^\prime$ and $t$ effects are constructive. They  can be
considerably relaxed for CPV phase $\phi_{sb}\la \pi/2$ and
$\phi_{sb}\ga 3\pi/2$. For $m_{t^\prime}=300$ GeV and
$\phi_{sb}=0$, $r_{sb}=0.07$ can still be tolerated. We see that
the bounds on $r_{sb}$ from $B\to X_s l^+ l^-$ are more
restrictive than those coming from $B\to X_s \gamma $.

Our bounds on $r_{sb}$ are similar to the finding of
Refs.~\cite{yanir,HNS}. Note that Ref.~\cite{yanir} does not use
$b\to s\gamma$ constraint and consider many others, and the bound
on $r_{sb}$ seems to come from $b\to X_s l^+ l^-$. We stress in
closing that $B_s$-$\bar{B}_s$ can constrain only slightly
$r_{sb}e^{i\phi_{sb}}$ \cite{AH2}, but rules out some region
around $\phi_{sb} \sim 0$ and large $r_{sb}$~\cite{HNS}, when
there is too much destruction between $t^\prime$ and $t$ effect.

\subsection{\boldmath $c\to u$, $s\to d$ and $b\to d$ transitions}

As seen in the previous sections, $B\to X_s\gamma$, $B\to
X_sl^+l^-$ and $B_s^0$--$\bar B_s^0$ mixing can constrain
$\lambda_{t'}=r_{sb}e^{i\phi_{sb}}$ for given $m_{t'}$. This would
have implications on $Z\to \bar sb + s\bar b$, as well as $t\to c$
transitions. To further constrain the $4\times 4$ CKM matrix
elements, one can consider other observables such as
$D^0$-$\bar{D}^0$ mixing and rare kaon decays $K_L\to \mu^+\mu^-$,
$K^+\to \pi^+\nu \bar{\nu}$, $\epsilon'/\epsilon$, and
$B_d$-$\bar{B}_d$ mixing and associated $CP$ violation.

Recent search for $D^0$-$\bar{D}^0$ mixing~\cite{PDG} puts an
upper bound on the mixing amplitude $|M_{12}^D| \leq 6.2\times
10^{-14}$ GeV at 95\% C.L. \cite{raz}. In this bound $CP$ violation
in the mixing has been included. Assuming that the long distance
contributions to the mixing amplitude are small, one can constrain
the fourth generation parameters, namely $m_{b'}$ and
$|V_{ub'}^*V_{cb'}|$. For fixed $m_{b'}$, and using analytic
formula from \cite{yanir,hatto}, we get a bound on
$|V_{ub'}^*V_{cb'}|$ as
\begin{eqnarray}
& &  |V_{ub'}^*V_{cb'}| \la \left\lbrace \begin{array}{ll}
7\times 10^{-3},  &  m_{b'}=240 \ {\rm GeV},\\
5.1\times 10^{-3},   &  m_{b'}=360 \ {\rm GeV}.
\end{array} \right.
\end{eqnarray}
The bound can be easily satisfied with a small $V_{ub^\prime}$,
which we would in general assume.

It has been demonstrated in \cite{yanir,HNS} that
$V_{t'd}^*V_{t's}$ is well constrained by ${\cal B}(K^+\to
\pi^+\nu\bar{\nu})$ $K_L\to \mu^+\mu^-$ and $\epsilon_K$. From
Ref.~\cite{HNS} and for $m_{t'}=300$ GeV, one can read that
$|V_{t's}^*V_{t'd}|$ up to $7\times 10^{-4}$ is tolerated both by
$\epsilon_K$ as well as by ${\cal B}(K^+\to \pi^+\nu\bar{\nu})$,
even for sizable $r_{sb} \sim 0.025$.

Overall, however, we will not be concerned with FCNC constraints
involving the first generation, except for discussions on $Z\to
\bar sb + s\bar b$ and $t\to c$ transitions. For FCNC transitions
involving $b^\prime$ and $t^\prime$, we will simply decouple the
first generation and not be concerned with the associated
constraints discussed here. The second generation, however, is
important, as we shall see.

%%%%%%%%%%%%%%%%%%%%%%%%%%%%%%%%%%%%%%%%%%%%%%%%%%%%%%%%%%%%%%%%
%%%%%%%%%%%%%%%%%%%%%%%%%%%%%%%%%%%%%%%%%%%%%%%%%%%%%%%%%%%%%%%%

\section{FCNC \boldmath $t\to cX$ Decays}

We now investigate FCNC involving heavy quarks with four
generation effect, which is the main focus of this paper. We start
with $t\to cZ, cH, cg, c\gamma$ decays in this section, turning to
$b^\prime\to bZ, bH, bg, b\gamma$ as well as $t^\prime\to tZ, tH,
tg, t\gamma$ in the next section.

To the best of our knowledge, the contribution from a sequential
fourth generation to top FCNC $t\to c Z$, $c H$, $c g$ and $c
\gamma $ has not yet been discussed.
The unitarity of the $4\times 4$ CKM matrix is expressed as
\begin{eqnarray}
\lambda_d + \lambda_s + \lambda_b + \lambda_{b^\prime}=0,
 \label{unit}
\end{eqnarray}
where $\lambda_f = V_{cf} V_{tf}^*$.
In our analysis we shall parameterize $\lambda_{b^\prime} =
V_{cb'}V_{tb'}^* \equiv r_{ct} e^{i \phi_{ct}}$.
Although naively the allowed range for $r_{ct}$ by unitarity is
rather large, however, the strength of $r_{ct}$ is correlated with
the strength of $r_{sb}$ by unitarity. In the standard
parameterization of $4\times 4$ CKM \cite{botella},
Ref.~\cite{HNS} has shown that for $m_{t^\prime}=300$ GeV, one
finds $|r_{ct}| = |\lambda_{b^\prime}| = |V_{cb'} V_{tb'}^*| \la
0.025$.
As well shall see below, for $t\to c$ transitions, $\lambda_d$,
$\lambda_s$, $\lambda_b$ turn out to be numerically irrelevant,
and only $\lambda_d+\lambda_s+\lambda_b = -\lambda_{b^\prime}$
matters. In what follows, to give an idea on the size of the
fourth generation effect on FCNC decays involving the top quark,
we will relax the $|r_{ct}| \la 0.025$ bound and take it to be in
the range of 0 to 0.05.

%%%%%%%%%%%%%%%%%%%%%%%%%%%%%%
\begin{figure}[t!]
\smallskip\smallskip
\vskip-.8cm
%%%%%%%%%%%%%%%%%%%%
\centerline{{
\epsfxsize3.4 in
\epsffile{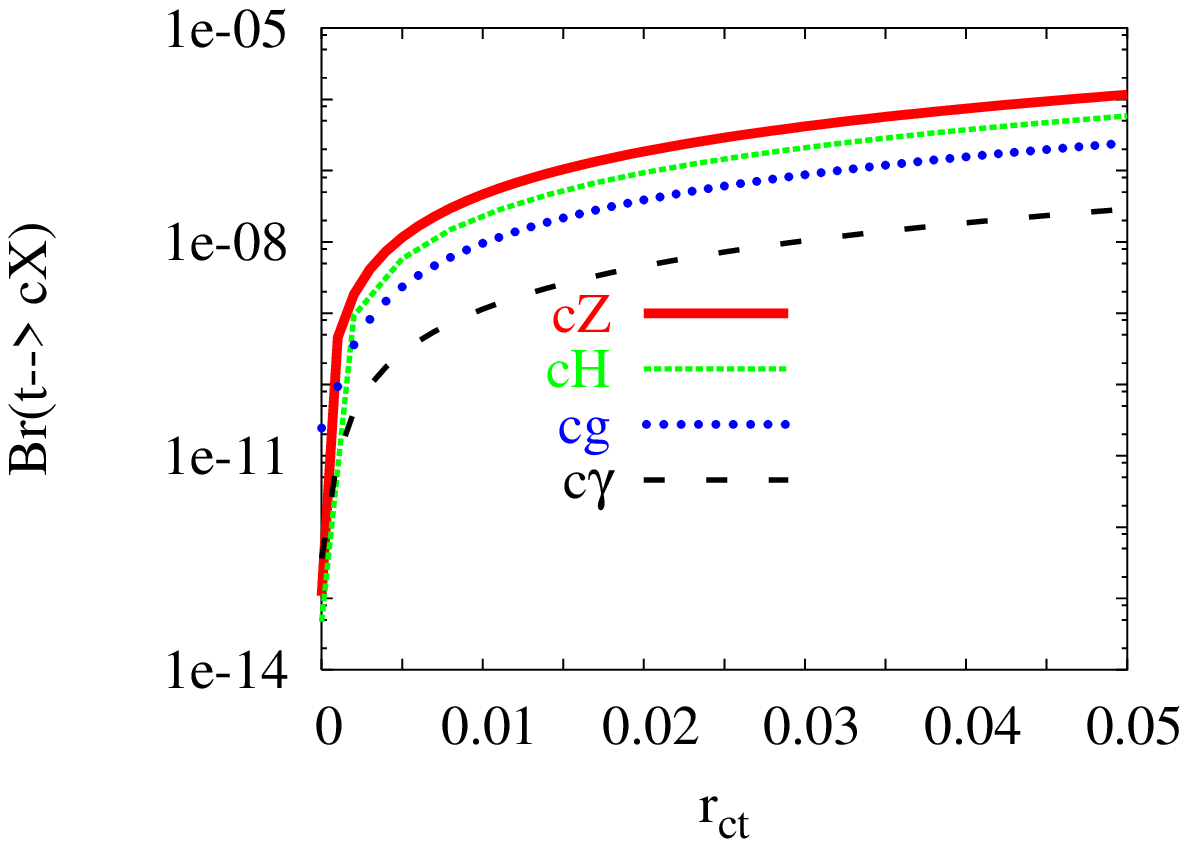}}
\hskip-1.4cm
\epsfxsize3.4 in
\epsffile{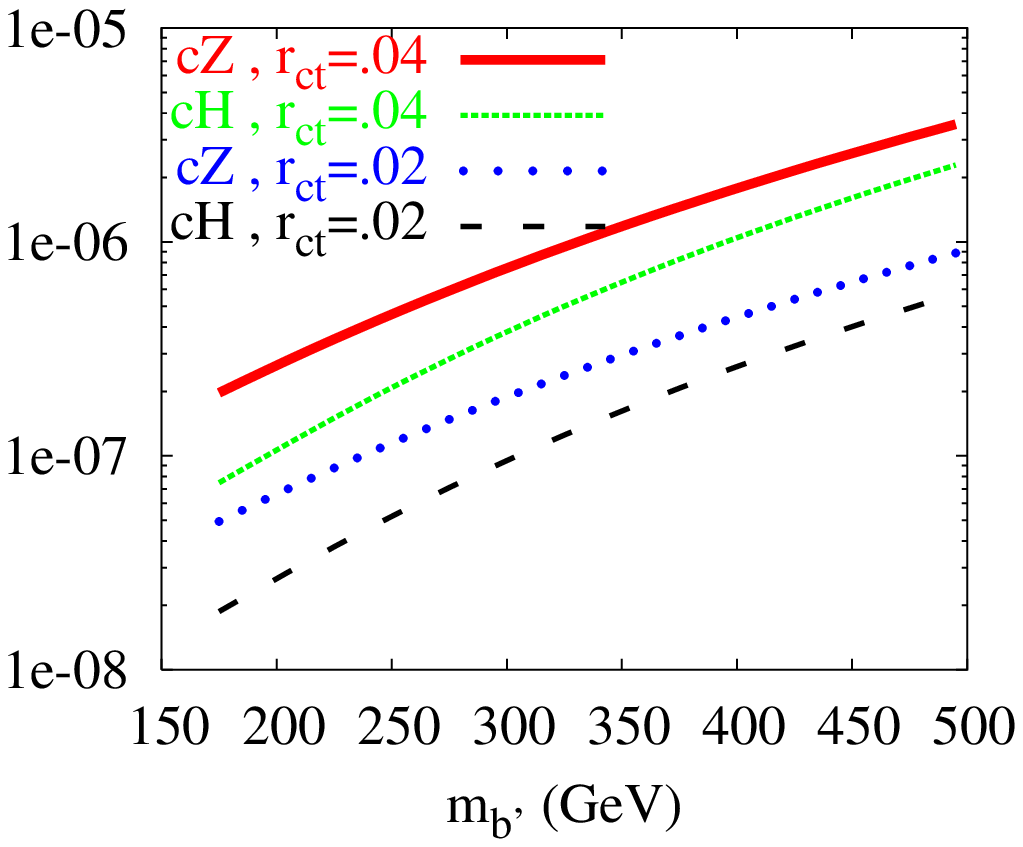} }
\smallskip\smallskip
\caption{Fourth generation effect on $t\to c\{Z, H, g, \gamma\}$
rate, with $M_H=115$ GeV, $m_{b^\prime}=300$ GeV (left) and
$r_{ct}=0.02$, 0.04  (right).}
\label{fig1}
\end{figure}

To discuss the numerical effects on all $t\to c$ transitions, we
fix our parameters as follows: $(m_t$, $m_b$, $m_c$, $m_s) =$
(174.3, 4.7, 1.5, 0.2) GeV, $\alpha^{-1} \approx 128$,
$\alpha_s(m_t) \approx 0.105$. The top width is taken as
$\Gamma_t=1.55$ GeV. All the computations of FCNC decay rates are
done with the help of the packages {FeynArts,
FormCalc}~\cite{FA2}, and with LoopTools and FF for numerical
evaluations~\cite{FF,LT}. The one-loop amplitudes are evaluated in
the 't Hooft--Feynman gauge using dimensional regularization.
Cross checks with well known SM FCNC processes are made, and
perfect agreement has been found. The numbers we show in
Eqs.~(\ref{eq1}) and (\ref{eq2}) for example, were obtained by
FormCalc \cite{FA2}.

We have check that the results for $t\to cX$ do not depend on
light quark masses $m_d$, $m_s$, $m_c$ and $m_b$ for any finite
$r_{ct}$, hence the one loop amplitudes ${\cal M}_{t\to c}$ can be
approximated as,
\begin{eqnarray}
{\cal M}_{t\to c}\propto r_{ct}e^{i\phi_{ct}}
[f(m_{b^\prime})-f(0)]\label{act},
\end{eqnarray}
where $f$ is some loop integral with implicit external quark mass
dependence, and light internal quark effects are summarized in
$f(0)$. It is clear from Eq.~(\ref{act}) that the CPV phase
$\phi_{ct}$ will not affect $t\to c$ transition rates at all.

In Fig.~\ref{fig1} we illustrate ${\cal B}(t \to \{cZ, cH, cg,
c\gamma \} )$ vs $r_{ct}$ for $m_{b^\prime}=300$ GeV (left plot),
and  vs $m_{b^\prime}$ (right plot) for $r_{ct}=0.02$ and 0.04.
$M_H=115$ GeV is assumed. The branching ratios are not sensitive
to the CPV phase $\phi_{ct}$ which has been dropped. As can be
seen from the left plot, we reproduce the SM results for ${\cal
B}(t \to \{cZ, cH, cg, c\gamma\})$ when $r_{ct}=0$ (where the
lighter quark masses are kept). As $r_{ct}$ increases, all
branching ratios can increase by orders of magnitude. For large
$r_{ct}=0.04$ and rather heavy $m_{b^\prime}\ga 350$ GeV,
branching ratios for $t \to cZ$ and $t\to cH$ can reach $\approx
10^{-6}$. But such large values of $r_{ct}$ and $m_{t^\prime}$,
combined, may run into difficulty with $b\to s$ transitions.

We see that $t\to c Z$ is much more enhanced compared to $t \to c
g$ and $t\to c\gamma$. This is due to the fact that the axial
coupling of the $Z$ boson is related to the unphysical Goldstone
boson, which is the partner of the physical Higgs boson before
symmetry breaking. Thus, the non-conserved part of the $Z$
coupling has a rather similar dependence on heavy internal quark
masses as the Higgs coupling, and both show nondecoupling of SM
heavy quarks in loop effects. For $t \to c g$ and $t\to c\gamma$,
they do not have this behavior because gauge invariance demands
conserved currents, and the heavy quark effects in the loop are
basically decoupled.

We conclude that fourth generation contributions can enhance top
FCNC couplings by orders of magnitude with respect to SM values,
reaching $10^{-7}$--$10^{-6}$ for $t \to cZ, cH$. But such
enhanced rates seem to be still out of experimental reach for the
foreseeable future. Nevertheless, search for FCNC $t\to c$ decays
should continue.

\section{FCNC \boldmath $b^\prime\to bX$ and $t^\prime\to tX$ transitions}

As mentioned above, EW precision measurements constrain
$|m_{t^\prime}-m_{b^\prime}|\la M_W$. This constraint does not
imply that $t^\prime$ is heavier than $b^\prime$. Therefore, we
shall address $b^\prime$ and $t^\prime$ decays for both $m_{b'} <
m_{t'}$ and $m_{b'} > m_{t'}$ situations.

Before describing our strategy for FCNC $b'\to b $ and $t'\to t $
decays, we review first the tree level decays of $b'$ and $t'$.
The possible Charged Current tree level decay modes are,
\begin{eqnarray}
& & b^\prime\to c W, t W^{(*)}, t^\prime W^*; \nonumber\\
& & t^\prime\to s W, b W, b^\prime W^*.
\end{eqnarray}
Due to expected suppression of the CKM elements $V_{ub'}$ and
$V_{t'd}$, the CC $b^\prime\to u W$ and $t^\prime\to d W$ decays
will be neglected in what follows. Since EW measurements constrain
the splitting between $t'$ and $b'$ to be less than about the $W$
mass, the decay $t'\to b' W^*$ or $b'\to t'W^*$ can occur only
with a virtual $W^*$ decaying as $W^*\to f_1f_2$.
For the tree decays $t^\prime, b^\prime \to \{b^\prime; t,
t^\prime\}W^*\to \{b^\prime; t, t^\prime\}f_1f_2$ involving
off-shell $W$ and heavy quark final state, we have used the
analytic expression from \cite{HR} and included all the light
fermion channels by using $\Gamma(Q'\to QW^*) = 9\Gamma(Q'\to Q
e\nu_e)$. In the following discussion of FCNC decays, we will
refrain from discussing the three body decays $b'\to b f\bar{f}$
and $t'\to t f\bar{f}$, with $f$ any light fermions or neutrinos.
These decays
could be comparable to $b' \to b\gamma$ or $t'\to t\gamma$
\cite{HSll}.

Turning to FCNC $b'\to b $ and $t'\to t $ transitions, we note
that FCNC $b^\prime$ decays have been extensively studied in the
literature \cite{AH,HR,HSll,hou1,sonibp,bigi}. Here, we update
those studies and investigate $t^\prime$ decays also, taking into
account recent experimental measurements. Motivated by the above
constraints on CKM matrix elements, in studying loop induced
$b'\to b $ and $t'\to t $ transitions, we will decouple the first
and second generations quarks. Therefore, the relevant CKM matrix
will be effectively $2\times 2$ and {\it real}. The one-loop
amplitudes take the following forms:
\begin{eqnarray}
& & {\cal M}_{b^\prime \to
  b} \propto r_{bb^\prime}[f(m_{t^\prime})-f(m_t)]\label{abbp},\\
& & {\cal M}_{t^\prime \to
  t} \propto r_{tt^\prime}[f(m_{b^\prime})-f(0)]\label{attp},
\end{eqnarray}
where $r_{bb'}=V_{t^\prime b}^*V_{t^\prime b^\prime}$, $r_{t
t^\prime}=V_{tb^\prime}V_{t^\prime b^\prime}^*$.
Note that the near degeneracy of light $m_d$, $m_s$ and $m_b$ at
the $t$, $b'$, $t'$ scale makes Eq. (\ref{attp}) a good
approximation without the assumption of neglecting the first two
generations.
Since
$V_{t^\prime b^\prime}\approx {\cal O}(1)$, $|r_{bb^\prime}|$ and
$|r_{t t^\prime}|$ are almost the same. From the discussion in
Section 2, we will take $r_{bb^\prime}$ in the range $0.05 -
0.25$. It is clear from Eqs.~(\ref{abbp}) and (\ref{attp}) that
the rates of FCNC $b^\prime \to b$ and $t^\prime \to t$ decays
will increase with $r_{bb^\prime}$. Like the $t\to c$ transition,
we expect that the mode $b^\prime \to b Z$ and $b^\prime \to b H$
will dominate over $b^\prime \to b \gamma$ and $b^\prime \to b g$.
We have checked numerically that, as long as
$|V_{cb'}|,|V_{t's}|\la 0.06$, assuming $2\times 2$ form in the
3rd and 4th generation sector and considering only $b'\to b$ and
$t'\to t$ transitions is a good approximation. This is because of
the lightness of the first 2 generations, as well as the expected
smallness of product of CKM elements. In the Appendix, we give a
comparison of the rates of suppressed $b' \to sX$, $t'\to cX$
decays with respect to $b'\to bX$, $t'\to tX$, which will depend
on the CKM elements $V_{t's}$ and $V_{cb'}$.

%%%%%%%%%%%%%%%%%%%%%%%%%%%%

%%%%%%%%%%%%%%%%%%%%%%%%%%%%%%
\begin{figure}[t!]
\smallskip\smallskip
%%%%%%%%%%%%%%%%%%%%%
\centerline{{ \epsfxsize3.4 in \epsffile{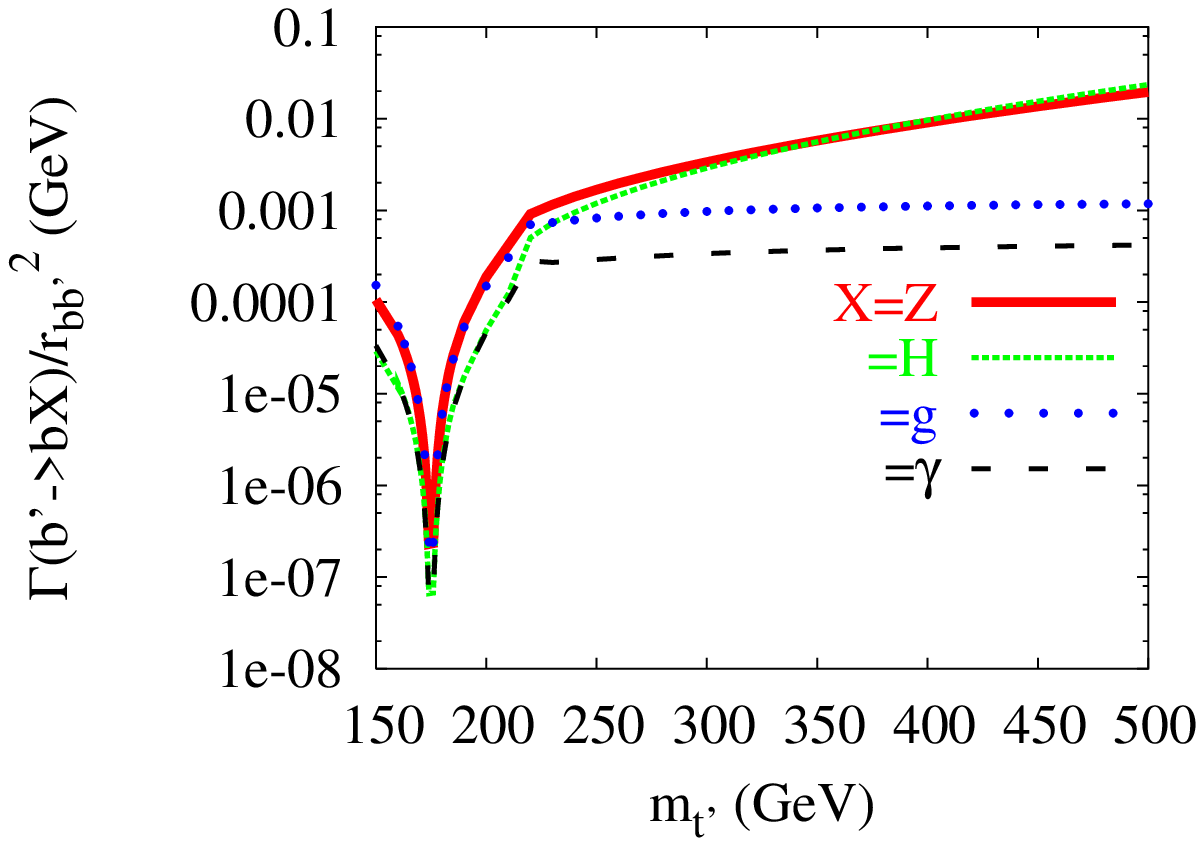}}
\hskip-1.6cm \epsfxsize3.4 in \epsffile{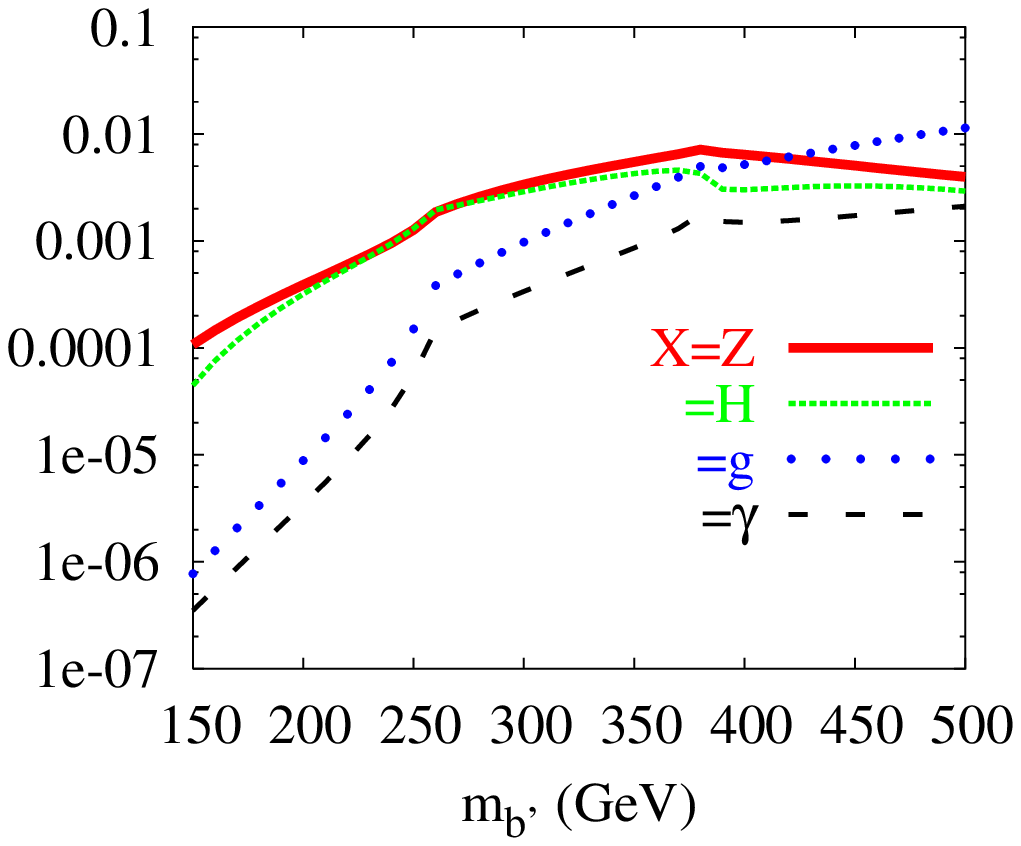} }
\smallskip\smallskip
\caption{Decay width of $\Gamma(b'\to b\{Z, H, g, \gamma\} )$
normalized to $r_{bb'}^2$ as a function of $m_{t'}$ for
$m_{b'}=300$ GeV (left), and as function of $m_{b'}$ for
$m_{t'}=300$ GeV (right).
}
\label{plot2}
\end{figure}
%%%%%%%%%%%%%%%%%%%%%%%%%%%%%%%%%%%%%%%%%

\subsection{Internal and External \boldmath $m_Q$ Dependence}

\begin{figure}[t!]
\smallskip\smallskip
%%%%%%%%%%%%%%%%%%%%%
\centerline{{ \epsfxsize3.4 in \epsffile{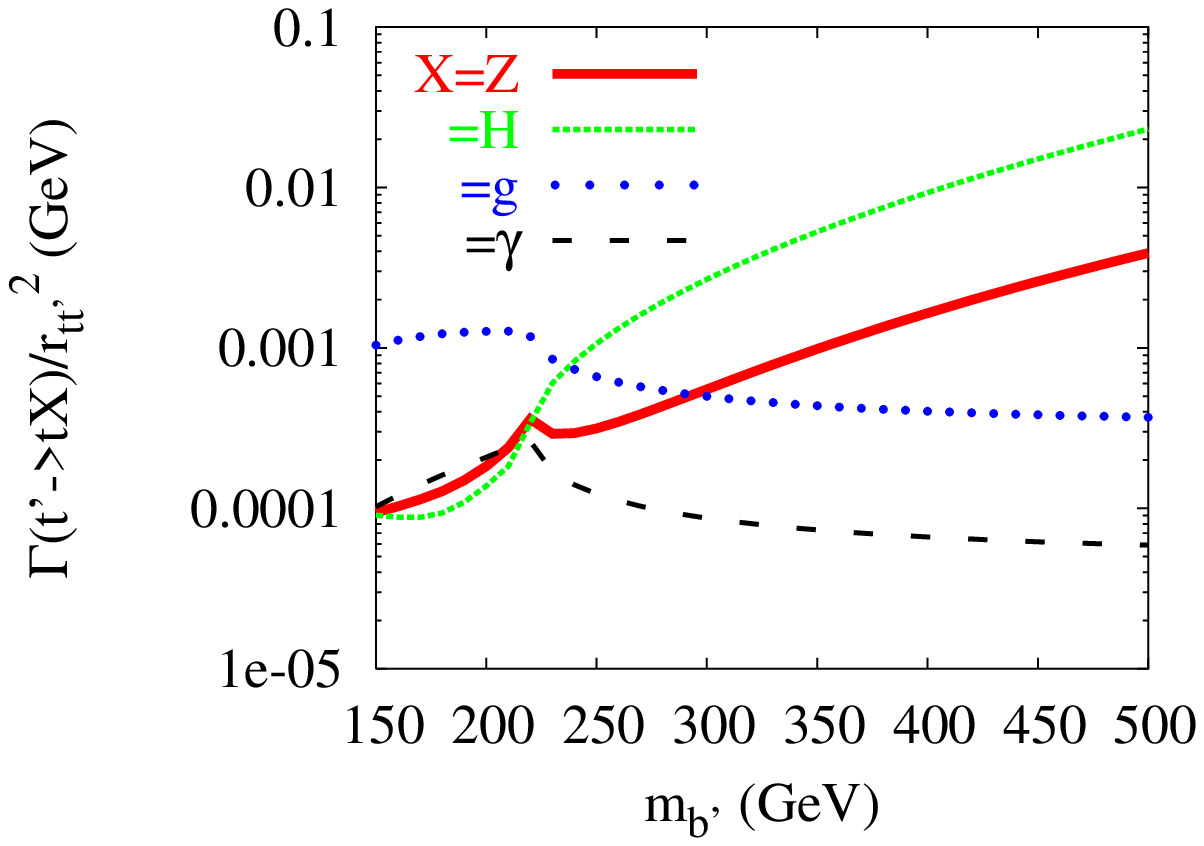}}
\hskip-1.6cm \epsfxsize3.4 in \epsffile{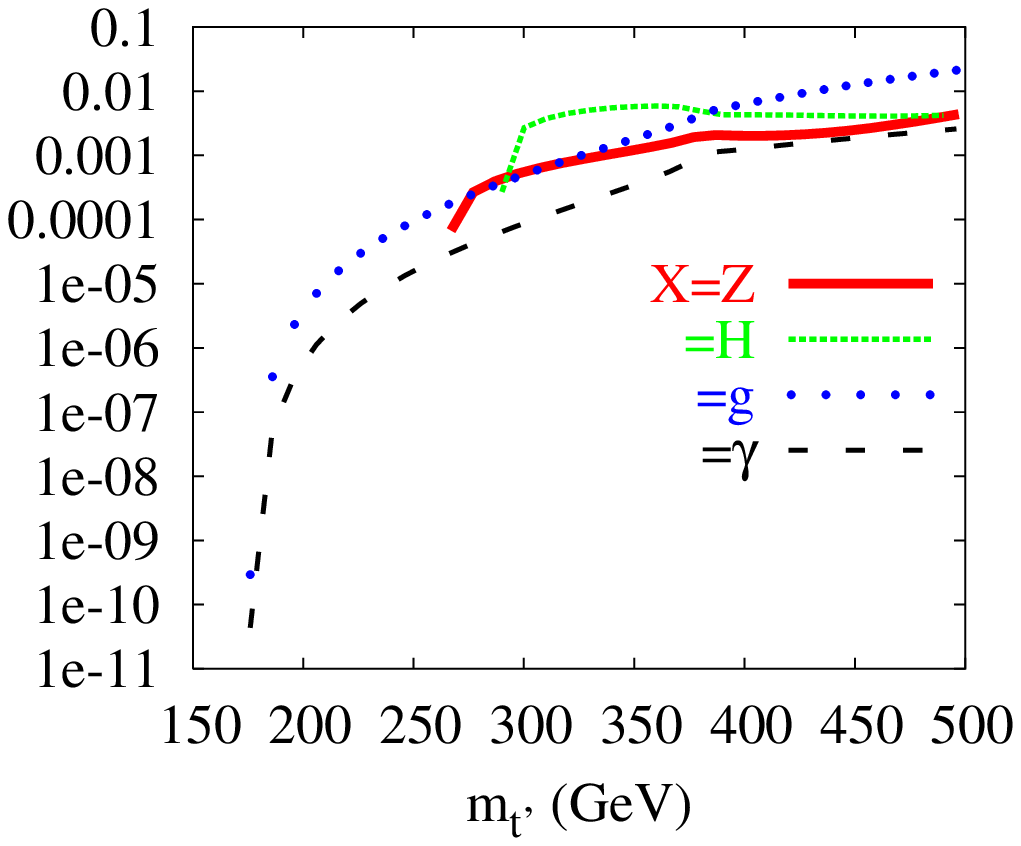} }
\smallskip\smallskip
\caption{Decay width of $\Gamma(t'\to t\{Z, H, g, \gamma\} )$
normalized to $r_{tt'}^2$ as a function of $m_{b'}$ for
$m_{t'}=300$ GeV (left), and as function of $m_{t'}$ for
$m_{b'}=300$ GeV (right).
} \label{plot3}
\end{figure}

Unlike the $t\to cX$ case, where there is one internal and one
external heavy quark each, with $m_t$ already fixed, $b'\to bX$
decay involves two internal and one external heavy quarks, while
$t'\to tX$ decay involves one internal and two external heavy
quarks. Before presenting our results for total width and
branching ratios, it is useful to discuss the sensitivity of the
FCNC $b'\to b$ and $t'\to t$ decay widths to both internal and
external heavy quark masses. To this end we give in
Figs.~\ref{plot2} and \ref{plot3} the decay widths $\Gamma(b'\to
bX)/r_{bb'}^2$ and $\Gamma(t'\to tX)/r_{tt'}^2$ for
$X=Z,H,g,\gamma$. For a fixed $m_{b'}$ (or fixed $m_{t'}$) value,
by the $\delta\rho$ constraint, the allowed range for $m_{t'}$ (or
$m_{b'}$) should be within $|m_{b'}-m_{t'}|\la M_W$. But for sake
of illustration, we will plot outside of such constraint.

In Fig.~\ref{plot2}(a), we show $\Gamma(b'\to bX)/r_{bb'}^2$ as
function of $m_{t'}$ in the 150--500 GeV range, with $m_{b'}$ held
fixed at 300 GeV. The big dip at $m_{t'} \sim m_t$ illustrates the
GIM cancellation of Eq. (\ref{abbp}), while the kink around
$m_{t'} \sim$ 220 GeV corresponds to $b'\to t'W$ cut. Above the
kink, the $b'\to bZ$ and $bH$ widths grow with increasing $m_{t'}$
because of the rising $t'$ Yukawa coupling. The $b'\to bg$ and
$b'\to b\gamma$ modes, however, stay almost constant. This is due
to decoupling of the $t'$ quark for large $m_{t'}$, and the effect
largely comes from the top. Thus, as one can see from the plot,
around $m_{t'} \sim 220$ GeV one has $\Gamma(b'\to bZ) \ga
\Gamma(b'\to bg) \ga \Gamma(b'\to bH) > \Gamma(b'\to b\gamma)$.
But for $m_{t'} \sim$ 500 GeV, one has $\Gamma(b'\to bH) \ga
\Gamma(b'\to bZ) > \Gamma(b'\to bg) > \Gamma(b'\to b\gamma)$, and
$\Gamma(b'\to b\{Z,H\})/r_{bb'}^2$ are slightly above
$0.01/r_{bb'}^2$ GeV.

In Fig.~\ref{plot2}(b), we plot $\Gamma(b'\to bX)/r_{bb'}^2$ as a
function of the decaying particle mass $m_{b'}$ for $m_{t'}$ fixed
at 300 GeV. Obviously, $\Gamma(b'\to bX)/r_{bb'}^2$ increase with
increasing $m_{b'}$ just from phase space. Above the $b'\to tW$
threshold at 255 GeV, the increase in rate with $m_{b'}$ slows,
more so for the $b'\to bZ$ and $bH$ modes. The $b'\to bg$ mode
becomes comparable to $b'\to bZ$ and $bH$ modes as the $b'\to t'W$
threshold of 380 GeV is approached. Passing this threshold,
interestingly, the $b'\to bZ$ and $bH$ widths start to decrease
slightly with $m_{b'}$, the external mass. The $b'\to bg, b\gamma$
modes, however, continue to rise with phase space, and $b'\to bg$
becomes the dominant mode above 400 GeV.

%%%%%%%%%%%%%%%%%%%%%%%%%%%%%%
%%%%%%%%%%%%%%%%%%%%%%%%%%%%%%%%%%%%%%%%%

We illustrate $\Gamma(t'\to tX)/r_{tt'}^2$ in Fig.~\ref{plot3} as a
function of $m_{b'}$ (left) and $m_{t'}$ (right). As one can see
from Fig.~\ref{plot3}(a), it is remarkable that $\Gamma(t'\to tg)$
is about one order of magnitude larger than $\Gamma(t'\to
t\{Z,H,\gamma\})$ for $m_{b'} < 220$ GeV, when $t'\to b'W$ is
open. Note that the gluon can only be radiated off the
$b^{(\prime)}$ quark, while $Z$, $H$ and $\gamma$ can also radiate
off the $W$ boson. For $m_{b'} > 220$ GeV, the $b'$ decouples from
$t'\to tg$, and $t\gamma$ and these widths decrease as $m_{b'}$
increase. But $\Gamma(t'\to t\{Z,H\})$ grow rapidly with
increasing $m_{b'}$. One can see also that $t'\to tH$ width
increases more rapidly with $m_{b'}$ as one passes through $b'W$
threshold, and remains considerably larger than $b'\to bZ$ for
large $m_{b'}$.

In Fig.~\ref{plot3}(b), we illustrate $\Gamma(t'\to tX)/r_{tt'}^2$
as function of $m_{t'}$, for $m_{b'}=300$ GeV held fixed. Of
course both $t'\to tZ$ and $t'\to tH$ are open only for
$m_{t'}>m_t+m_Z$ and $m_{t'}>m_t+M_H$ respectively. Once the Higgs
decay mode is open it dominate over $Z$ and gluon modes. But when
$m_{t'}$ crosses $380$ GeV, the $t'\to b'W$ threshold, the gluon
mode starts to dominate over the other three decay modes. This
behavior is similar to what is seen in Fig.~\ref{plot3}(a) for
lower $b'$ and $t'$ masses.

The FCNC widths considered in the following subsections are
special cases of those presented in Figs.~\ref{plot2} and
\ref{plot3}.

%%%%%%%%%%%%%%%%%%%%%%%%

\subsection{\boldmath $b'$ and $t'$ Widths}

For illustration, we fix $m_{t^\prime}=300$ GeV and consider the
cases of $m_{b^\prime} = 240$ GeV to illustrate $m_{b^\prime} <
m_{t^\prime}$, and $m_{b^\prime}=360$ GeV to illustrate
$m_{b^\prime} > m_{t^\prime}$. We shall use the examples of
$|V_{cb'}|\approx |V_{t^\prime s}|\approx V_{cb}\approx 0.04$ and
$|V_{cb'}|\approx |V_{t^\prime s}|=10^{-3}$ to illustrate sizable
versus suppressed $V_{cb'}$. The results for intermediate strength
$|V_{cb'}|\approx |V_{t^\prime s}|$ can be inferred from our plots
given in the next two subsections.

Let us first give the estimates of different decay widths that
contribute to $b^\prime$ and $t^\prime$ decays. For $b'$, the FCNC
and CC tree level decay widths are given (in GeV) as:
\begin{eqnarray}
\Gamma(b^\prime\to bZ, bH,bg, b\gamma) & = & \left\lbrace
\begin{array}{ll}
 (0.98,0.95,0.07,0.028)\times 10^{-3} r_{bb'}^2, & m_{b'}=240\ {\rm GeV}, \\
 (0.61,0.45,0.33,0.11)\times 10^{-2} r_{bb'}^2, &  m_{b'}=360\ {\rm GeV},
\end{array} \right. \\
\Gamma(b^\prime\to cW, tW^{(*)}, t^\prime W^*) & = & \left\lbrace
\begin{array}{ll}
 4.38 |V_{cb'}|^2, \ 0.0046
 |V_{tb'}|^2, \ 0, & m_{b'}=240\ {\rm GeV}, \\
 15.2 |V_{cb'}|^2, \ 6.66
 |V_{tb'}|^2, \ 0.0031, & m_{b'}=360\ {\rm  GeV},
\end{array} \right.
\end{eqnarray}
where $V_{t'b'} \simeq 1$ is assumed. The total decay width is
then (in GeV)
\begin{eqnarray}
& & \Gamma_{b^\prime} = \left\lbrace \begin{array}{ll}
0.007 r_{bb'}^2 + 4.38 |V_{cb'}|^2, & \ \ \ \ m_{b'}=240 \ {\rm GeV}, \\
 6.68 r_{bb'}^2 + 15.2 |V_{cb'}|^2 + 0.0031,
  & \ \ \ \ m_{b'}=360 \ {\rm GeV}. \label{totb}
\end{array} \right.
\end{eqnarray}
Note that we shall take $|V_{tb'}|$ for tree and
$|V_{tb}^*V_{tb'}| \simeq |V_{t'b'}V_{tb'}^*| \simeq |r_{bb'}|$
for loop as the same strength in our plots for simplicity.
We have thus combined the $b'\to tW^{(*)}$ and $b'\to bX$ widths
in Eq. (\ref{totb}).

For $t'$, one has (in GeV)
\begin{eqnarray}
\Gamma(t^\prime\to tZ, tH,tg, t\gamma) & = & \left\lbrace
\begin{array}{ll}
(0.29,0.83,0.74,0.14)\times 10^{-3} r_{tt'}^2, & m_{b'}=240 \ {\rm GeV}, \\
 (1.1,6.1,0.43,0.07) \times10^{-3} r_{tt'}^2, &  m_{b'}=360 \ {\rm GeV},
\end{array} \right. \\
\Gamma(t^\prime\to bW, sW, b^\prime W^*) & = & \left\lbrace
\begin{array}{ll}
8.72 |V_{t'b}|^2, \ 8.73 |V_{t's}|^2, \ 0.003,
 & \ \ \ \  m_{b'}=240 \ {\rm GeV}, \\
8.72 |V_{t'b}|^2, \ 8.73 |V_{t's}|^2, \ 0, & \ \ \ \ m_{b'}=360 \
{\rm GeV},
\end{array} \right.
\end{eqnarray}
where $V_{t'b'} \simeq 1$ is assumed.
The total decay width is (in GeV)
\begin{eqnarray}
\Gamma_{t^\prime} = \left\lbrace \begin{array}{ll}
8.72 r_{tt'}^2 + 8.73 |V_{t's}|^2 + 0.003, & \ \ \ \
m_{b'}=240 \ {\rm  GeV}, \\
 8.73 r_{tt'}^2 + 8.73 |V_{t's}|^2,
 & \ \ \ \ m_{b'}=360 \ {\rm GeV}. \label{tott}
\end{array} \right.
\end{eqnarray}
Again we shall take $|V_{t'b}| \simeq |V_{t'b}^*V_{tb}| \simeq
|V_{t'b'}^*V_{tb'}| \simeq |r_{tt'}|$ in our plots for simplicity.
So $t'\to bW^{(*)}$ and $t'\to tX$ widths are combined in
Eq.~(\ref{tott}).

It is clear from the above equations that in general the CC decays
$b'\to \{cW, tW\}$ and $t'\to \{bW,sW\}$ would dominate $b'$ and
$t'$ rates, respectively. One may think that the $b^\prime\to c W$
and $t^\prime\to s W$ decays should be subdominant since
$|V_{t^\prime s}|^2\approx |V_{cb^\prime }|^2 \ll |V_{t^\prime
b}|^2\approx |V_{tb^\prime }|^2$ seem plausible. However, in the
plausible HNS scenario \cite{HNS} which we will discuss later,
$|V_{t^\prime s}|^2\approx |V_{cb^\prime }|^2$ is not much smaller
than $|V_{t^\prime b}|^2\approx |V_{tb^\prime }|^2$, and the
$b^\prime\to c W$ and $t^\prime\to s W$ decays can compete with
$b^\prime\to t W$ and $t^\prime\to b W$, and could even dominate.
For the lighter $m_{b'} = 240$ GeV case, if $|V_{cb'}| \ll
r_{bb'}$ then $b'\to cW$ is CKM suppressed, and FCNC $b'\to bZ$,
$bH$ are comparable to $b'\to tW^*$. The latter could be further
suppressed if $m_{b'} < 240$ GeV. But depending on the level of
suppression for $|V_{c b^\prime}|$, $b^\prime \to c W$ could also
be comparable with the loop-induced FCNC $b^\prime \to bZ$ and
$b^\prime \to bH$ decays. Such a scenario has been studied in
Ref.~\cite{AH} with $m_{b'}\la 200$ GeV and where it has been
assumed that $|V_{c b^\prime }/(V_{t^\prime b^\prime}
V_{bt^\prime}^*)|\approx 10^{-3}$. It is still relevant at the
Tevatron.
%

%%%%%%%%%%%%%%%%%%%%%%%%%%%%%%
\begin{figure}[t!]
\smallskip\smallskip
%%%%%%%%%%%%%%%%%%%%%
\centerline{{ \epsfxsize3.4 in
\epsffile{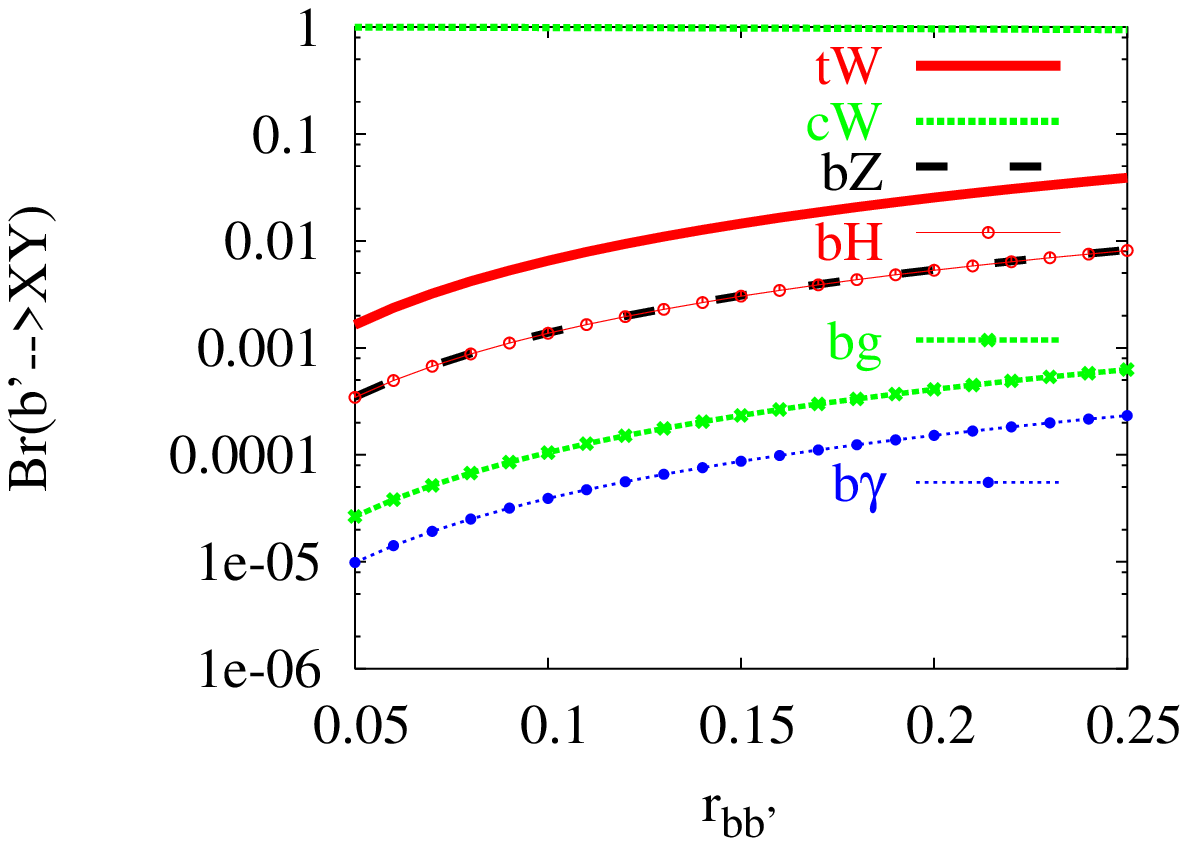}}
\hskip-1.6cm
\epsfxsize3.4 in
\epsffile{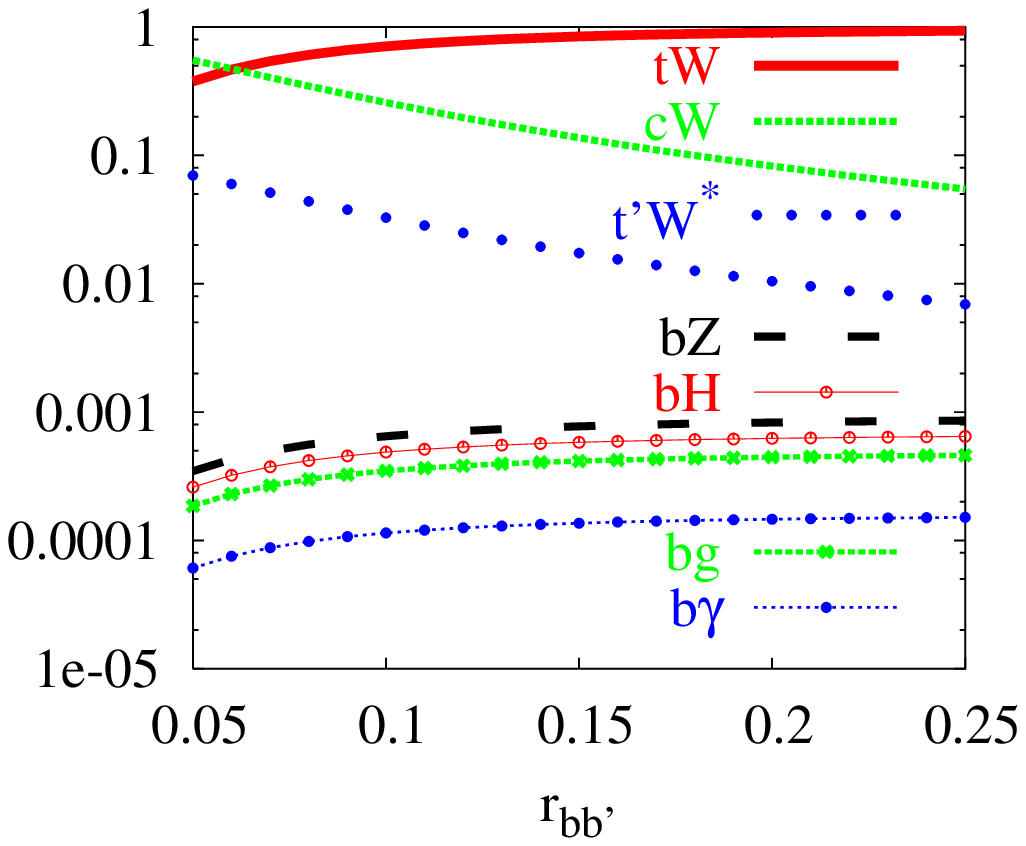}}
\smallskip\smallskip
\caption{Branching ratios for $b^\prime$ vs $r_{bb^\prime}$ for
$m_{b^\prime}=240$ GeV (left), 360 GeV (right), with
$m_{t^\prime}=300$ GeV and $|V_{cb^\prime}|=0.04$.}
\label{plot4}
\end{figure}
%%%%%%%%%%%%%%%%%%%%%%%%%%%%%%%%%%%%%%%%%

\subsection{Phenomenology of \boldmath $b'$ Decay}

We show in Fig.~\ref{plot4}(a) the various $b'$ decay branching
ratios for $|V_{cb'}| = 0.04 \sim V_{cb}$. In this case, the
$b^\prime\to cW$ process is the dominant decay mode for
$m_{b^\prime}=240$ GeV, which illustrates $m_{b^\prime} <
m_{t^\prime}$. Thus, one should search for $b'\bar b'$ via $c\bar
c W^+W^-$. The loss of $b$-tagging as a powerful tool will make
this study more challenging.
The off-shell W decay $b^\prime\to tW^*$ is open but is in the
range of $10^{-3}$--$10^{-2}$.
The size of the loop-induced FCNC $b^\prime\to bZ, bH$ decays is
of order $10^{-3}$, but could be larger for a lighter $b'$. Such
strength for FCNC is sizable when compared with $t\to cX$.
According to Eq.~(3), top FCNC of the order $10^{-5}$ can be
measured at LHC or ILC. For heavier quarks such as $b'$ and $t'$,
the production cross sections are smaller than $t\bar{t}$ case,
resulting in a smaller number of events. We therefore expect that
the sensitivity to heavy quark FCNC decays will be less than what
we have listed in Eq.~(3).
%Eq.~(\ref{LHC}) are readily measurable at the LHC. Even the $bg$
%and $b\gamma$ modes, at $10^{-4}$ order, can be probed.
Still, it should be promising to probe heavy quark FCNC decays up
to the order $10^{-3}$ at the LHC or ILC. A higher luminosity run
and/or a higher energy machine would improve the sensitivity.

The results for $m_{b^\prime} = 360 > m_{t^\prime}=300$ GeV is
given in Fig.~\ref{plot4}(b), where the $b^\prime$ decay channels
are richer. The $b^\prime\to t^\prime W^*$ mode is now open but
still subdominant. Since $m_{b^\prime}$ is higher, the decay
$b^\prime\to t W$ is enhanced to over 30\% for small
$r_{bb^\prime}\approx 0.05$, and approaches 100\% for large
$r_{bb^\prime}$ approaching ${\cal O}(\lambda)$. For
$r_{bb^\prime}\ga 0.06$ and $|V_{cb^\prime}|=0.04$, $b^\prime\to
tW$ dominates over $b'\to cW$ mode. $|V_{cb^\prime}| > 0.04$ would
delay the dominance of $tW^*$ over $cW$, but for $|V_{cb^\prime}|
< 0.04$, the $b^\prime\to cW$ mode will be further suppressed. The
search strategy should be via $b'\bar b' \to t\bar t W^+W^- \to
b\bar bW^+W^-W^+W^-$, which has 4 $W$ bosons. However, depending
on strength of $b'\to cW$, one could have $t\bar c W^+W^-$, $c\bar
c W^+W^-$, resulting in 3 or 2 $W$ bosons only, with reduced
$b$-tagging discrimination. The $t'W^*$ mode would pose a further
challenge with off-shell $W$s.
For the FCNC decays, note that $b'\to bg$ is close to $bZ$ and
$bH$ rate of the  order $10^{-3}$, while $b'\to b\gamma$ is at
$10^{-4}$ order. The LHC should be able to probe a major part of
this range of rates.

%%%%%%%%%%%%%%%%%%%%%%%%%%%%%%
\begin{figure}[t!]
\smallskip\smallskip
%%%%%%%%%%%%%%%%%%%%%
\centerline{{ \epsfxsize3.4 in \epsffile{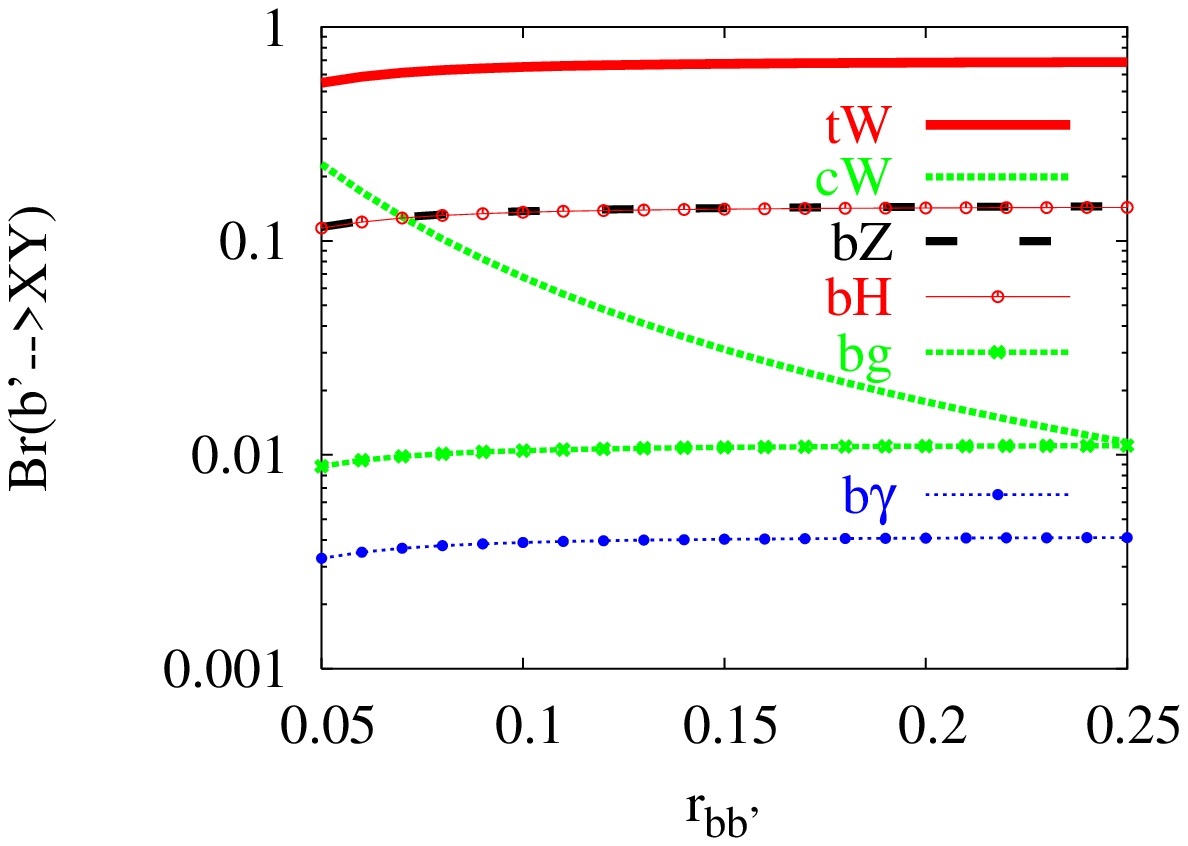}}  \hskip-1.6cm
\epsfxsize3.4 in \epsffile{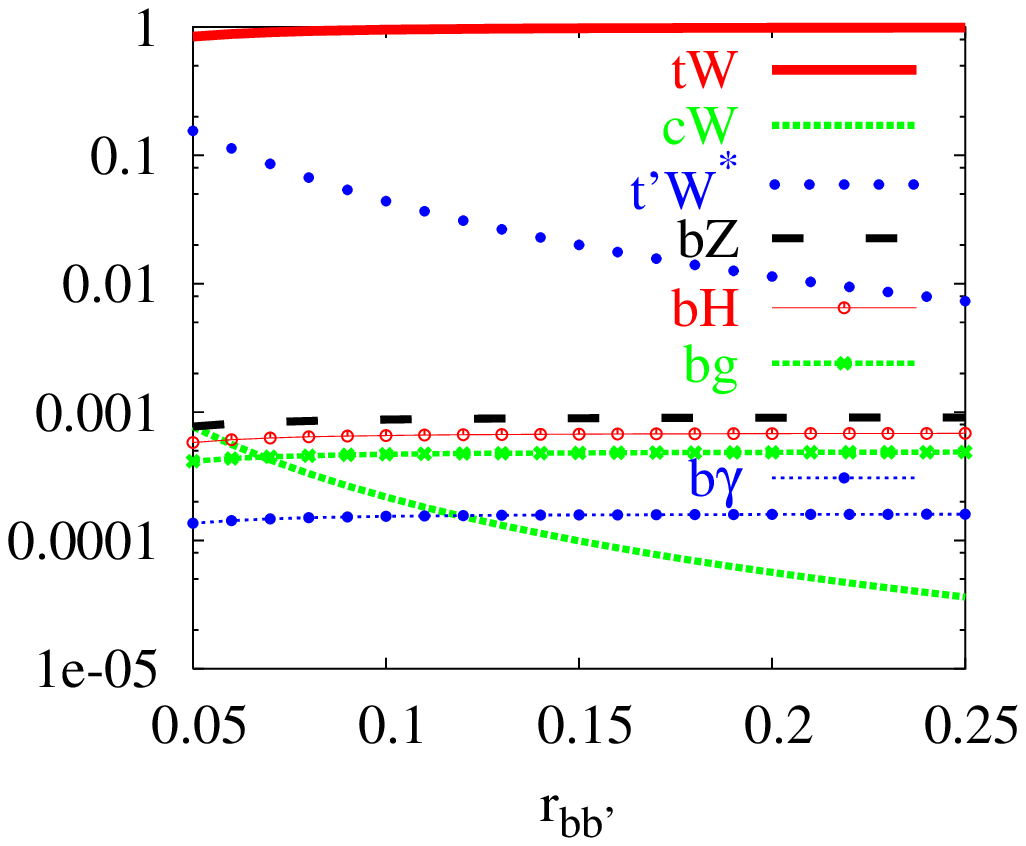} }
\smallskip\smallskip
\caption{Same as Fig.~\ref{plot4} with $|V_{cb^\prime}|=10^{-3}$.}
\label{plot5}
\end{figure}
%%%%%%%%%%%%%%%%%%%%%%%%%%%%%%%%%%%%%%%%%

For $|V_{cb^\prime}|$ as small as $10^{-3}$, as can be seen from
Fig.~\ref{plot5}(a), for $m_{b^\prime}=240$ GeV and
$m_{t^\prime}=300$ GeV, $b'\to tW^*$ is the dominant mode for
$r_{bb'}$ in range of 0.05--0.25,
with $bZ$ and $bH$ comparable. The study of $b'\bar b' \to t\bar t
W^*W^*$ should be undertaken. The off-shell nature of the $W$
would make it somewhat more troublesome. Since $b'\to bZ$, $bH$
could easily be a few 10\% (e.g. a slightly lighter $b'$), one
should really be searching for $t\bar t W^*W^*$, $t\bar b W^*Z$,
$t\bar b W^*H$, $b\bar b ZH$ simultaneously, which is a rewarding
if not complicated program.
Furthermore, since $b'\to tW^*$ is highly sensitive to $m_{b'}$,
and could be suppressed by smaller $m_{b'} < 240$ GeV, the FCNC
$b' \to bZ, bH$ decays could still dominate for relatively light
$m_{b'}$ just above 200 GeV. This could help uncover the Higgs
boson \cite{AH}!
For $r_{bb'}\approx 0.05$, $b'\to cW$ is just below $b'\to bZ$ and
$bH$. Its branching ratio decreases for larger $r_{bb'}$, becoming
comparable in size with $b'\to bg > b\gamma$ at the $10^{-2}$
level as $r_{bb'}$ approaches 0.2. The $b'\to b\gamma$ mode is at
a few $\times 10^{-3}$.
Thus, this scenario of relatively light $b'$ and very suppressed
$V_{cb'}$ is the most interesting one for FCNC $b'$ decays.

For $m_{b^\prime}=360$ GeV and $m_{t^\prime}=300$ GeV, as seen in
Fig.~\ref{plot5}(b),
$b'\to tW$ is now fully open and is of order 100\% for the full
range of $r_{bb'}$. It is followed by $b' \to t^\prime W^*$ with
rates in the range of $(1 - 10)\%$, since $V_{t'b'} \cong 1$,
dropping as $r_{bb'}$ increases.
In this case of heavy $m_{b'}$, the FCNC $b'\to bZ$, $bH$ and $bg$
decays are comparable and just below $10^{-3}$, with $b'\to
b\gamma$ just below at a few $\times 10^{-4}$ in rate. Compared to
$t\to cX$, even to $b\to sX$, these are still rather sizable rates
for FCNCs, and in view of Eq. (\ref{LHC}), they can be probed at
the LHC if the heavy quarks are not too heavy. The FCNCs are
dominant over CC $b'\to cW$ decay for $r_{bb'}\ga 0.1$, which is
at the $10^{-4}$ order or less.

There is, therefore, a rather broad range of possibilities for
$b'$ decay, depending on $m_{b'}$, $V_{cb'}$ and $V_{tb'}$. For
$m_{b'} \la 240$ GeV and very small $V_{cb'}$, FCNC dominance is
possible~\cite{AH}. For $m_{b'} > m_t + M_W$, the dominance of
$b'\to tW$ implies $b'\bar b' \to t\bar tW^+W^- \to b\bar
bW^+W^+W^-W^-$, or 4$W$s plus 2 $b$-jets, which should be of
interest at LHC. In between, the signal varies in richness and
complexity, but the FCNC are always within reach at the LHC.

%%%%%%%%%%%%%%%%%%%%%%%%%%%%%%
\begin{figure}[t!]
\smallskip\smallskip
%%%%%%%%%%%%%%%%%%%%%
\centerline{{ \epsfxsize3.4 in \epsffile{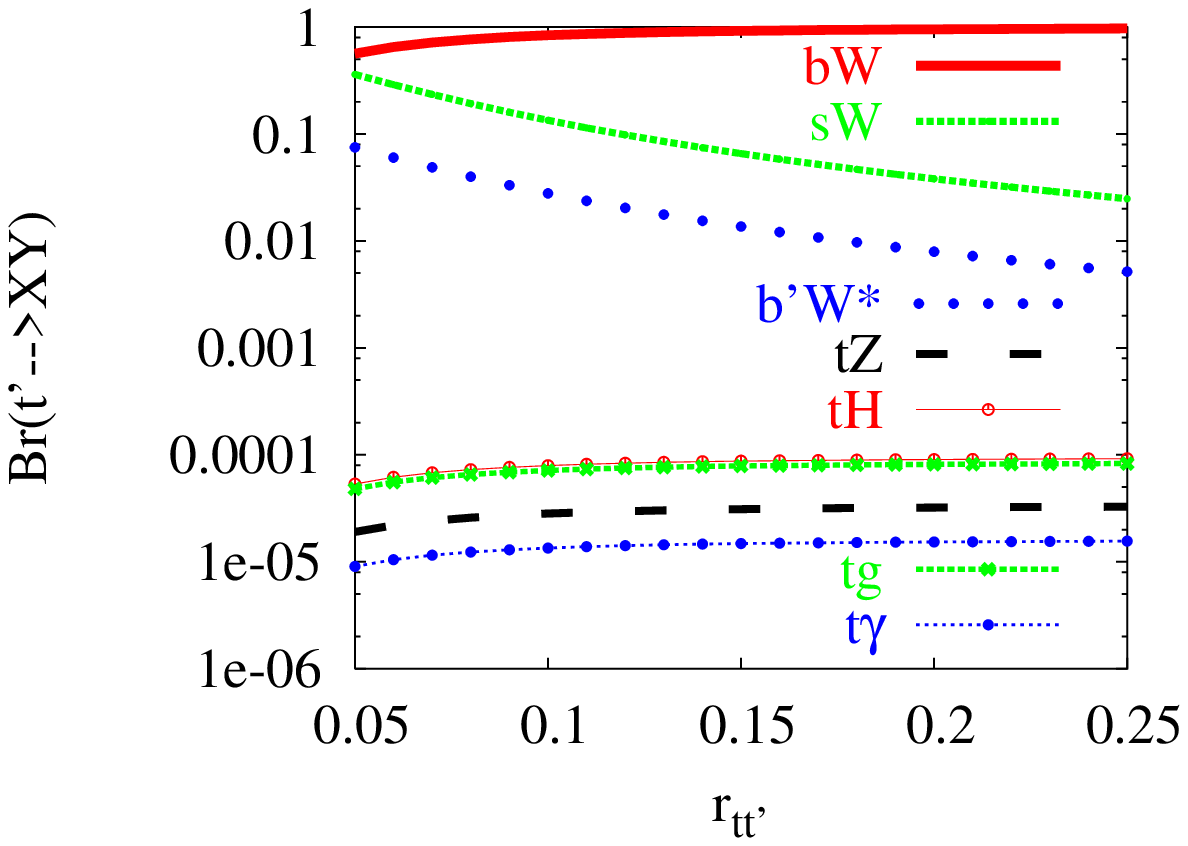}}  \hskip-1.6cm
\epsfxsize3.4 in \epsffile{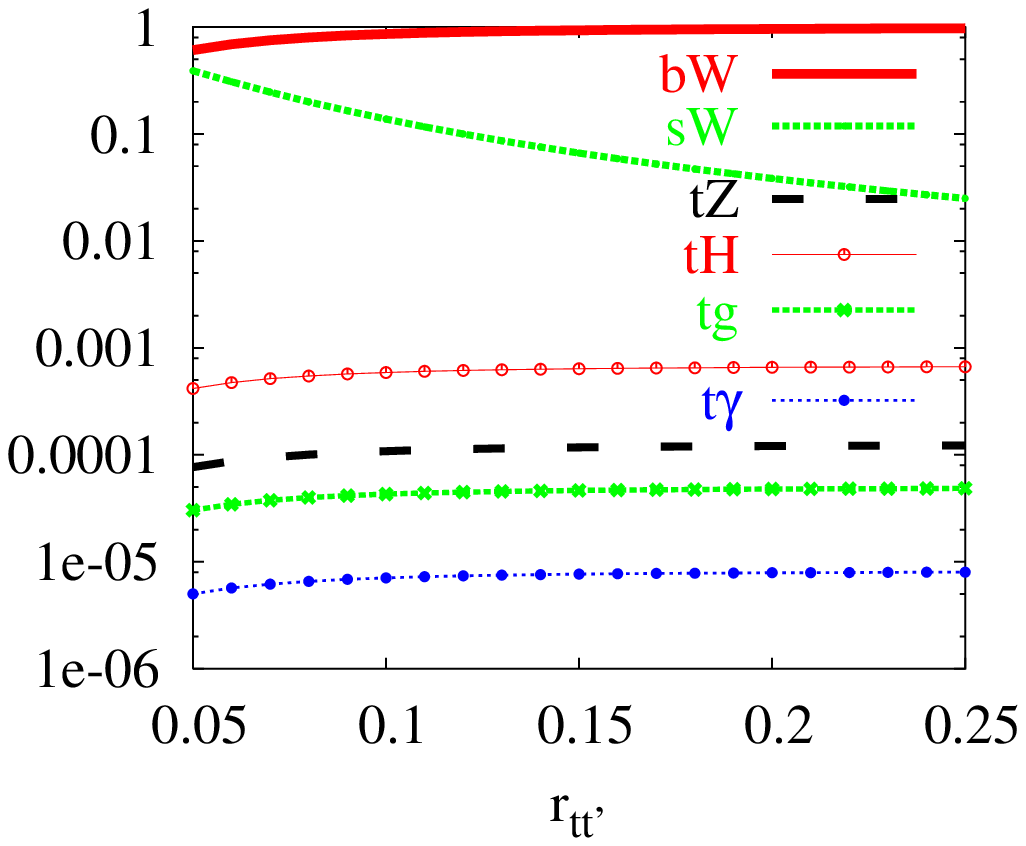} }
\smallskip\smallskip
\caption{Branching ratio for $t^\prime$ vs
  $r_{tt^\prime}$ for $m_{b^\prime}=240$ GeV (left),
  360 GeV (right), with $m_{t^\prime}=300$, and
  $|V_{cb^\prime}|=|V_{t^\prime s}|=0.04$.} \label{plot6}
\end{figure}
%%%%%%%%%%%%%%%%%%%%%%%%%%%%%%%%%%%%%%%%%

\subsection{Phenomenology of \boldmath $t'$ Decay}

\begin{figure}[t!]
\smallskip\smallskip
%%%%%%%%%%%%%%%%%%%%%
\centerline{{ \epsfxsize3.4 in \epsffile{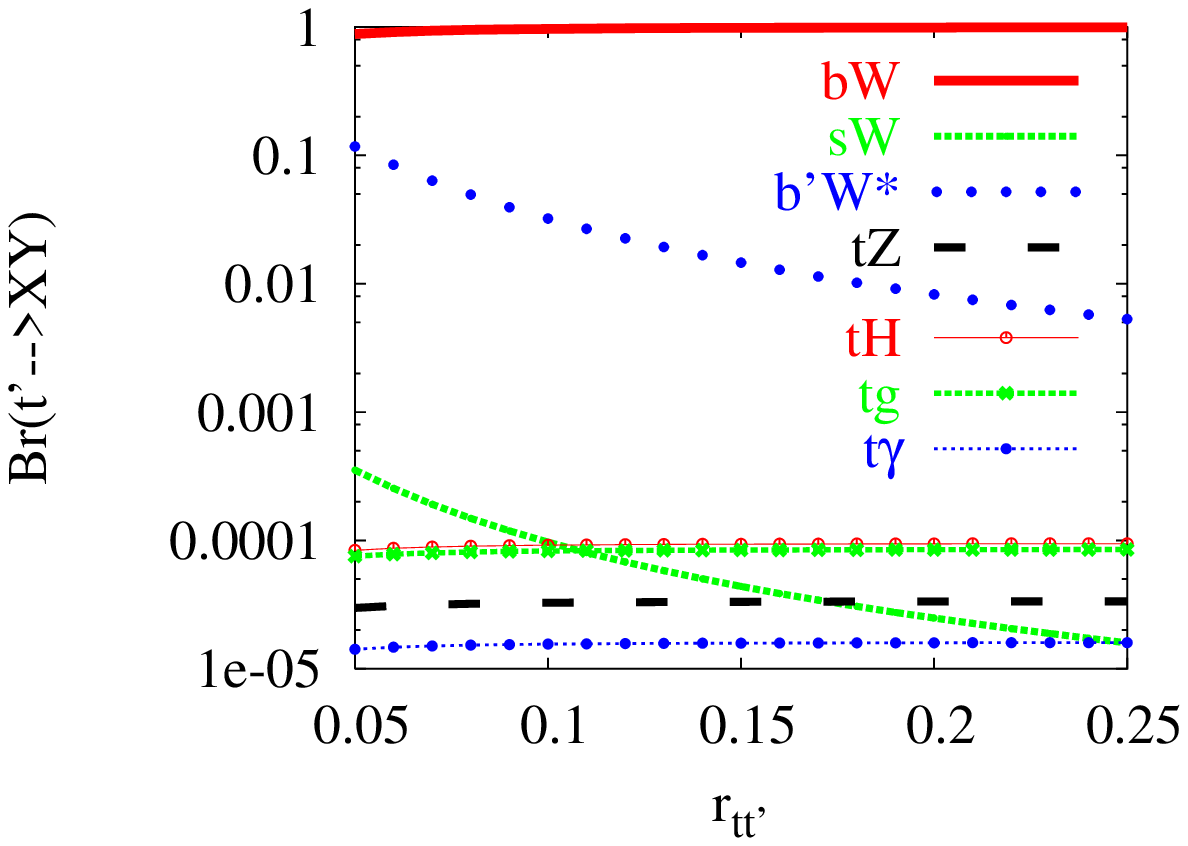}}  \hskip-1.6cm
\epsfxsize3.4 in \epsffile{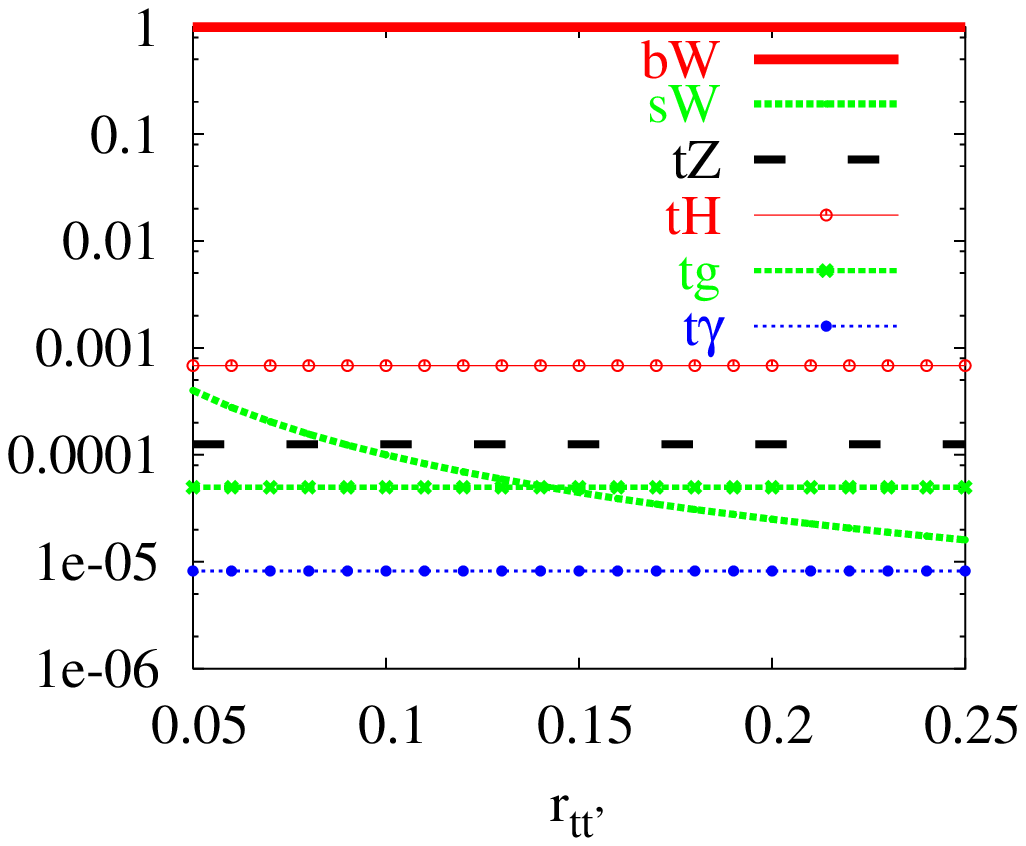} }
\smallskip\smallskip
\caption{Same as Fig.~\ref{plot6} with
$|V_{cb^\prime}|=|V_{t^\prime s}|=10^{-3}$.}
\label{plot7}
\end{figure}

Let us turn now to $t^\prime$ decays.
As shown in Figs.~\ref{plot6} and \ref{plot7} for $m_{t'} = 300$
GeV, $t^\prime\to bW$ is fully open and dominates over $t'\to sW$
for $|V_{t^\prime s}|\approx |V_{cb^\prime}| \approx 0.04$.

For $m_{t^\prime} > m_{b^\prime}$, as illustrated by
Fig.~\ref{plot6}(a) for $m_{b^\prime} = 240$ GeV, the decay mode
$t^\prime \to b^\prime W^*$ is open but kinematically suppressed,
and its branching ratio is in the range of $10^{-2}$--$10^{-1}$.
In the case of enhanced  $|V_{t^\prime s}| \ga 0.04$, $t^\prime
\to s W$ could in principle dominate over $t^\prime \to b W$ mode.
For the FCNC decays, note that $t'\to tH$ and $tg$ are comparable
at $10^{-4}$ order, with $tZ$ slightly below, followed by
$t\gamma$ around $10^{-5}$ order. Though more difficult than $b'$
case, these rates are above the $t\to cX$ rates, and may be
measurable at the LHC.

For $m_{t^\prime} < m_{b^\prime}$ as illustrated by
Fig.~\ref{plot6}(b), the $t'\to b'W^*$ decay is forbidden.
With the heavier $b'$, the $t'\to tH$ rate is raised to close to
$10^{-3}$, followed by $t'\to tZ$, which is slightly above the
$t'\to tg$ rate around $10^{-4}$ order. The $t'\to t\gamma$ mode
is around $10^{-5}$. The $t'\to tH$ rate may be measurable at the
LHC, but the other rates may be more difficult.

%%%%%%%%%%%%%%%%%%%%%%%%%%%%%%
%%%%%%%%%%%%%%%%%%%%%%%%%%%%%%%%%%%%%%%%%

For suppressed $|V_{t^\prime s}|\ll 0.04$, e.g. $|V_{t^\prime
s}|\approx 10^{-3}$, $t'\to sW$ becomes suppressed. For
$m_{b^\prime}=240$ GeV case shown in Fig.~\ref{plot7}(a), $t'\to
sW$ mode is of comparable size to the FCNC decay modes for
$r_{tt'}\la 0.08$, and drops lower for larger $r_{tt'}$. The other
features, including FCNC, are not very sensitive to $V_{t's}$ and
similar to Fig.~\ref{plot6}(a).
For $m_{b^\prime}=360$ GeV $ > m_{t'}=300$ GeV, as shown in
Fig.~\ref{plot7}(b), $t'\to b'W^*$ is forbidden, while the
suppressed $t^\prime \to sW$ mode drops below $t^\prime \to tZ$
for $r_{tt'}\la 0.1$. Otherwise the features are similar to
Fig.~\ref{plot6}(b).

Thus, the $t'$ quark behaves like a heavy top quark, with $t' \to
bW$ the dominant decay mode. But the decay modes are still rather
rich with $t'\to b'W^*$ and $sW$ possibilities.
Unlike the top, FCNC $t'\to tH$, $tZ$, $tg$ decays are around
$10^{-4}$, with $t'\to tH$ reaching branching ratio of the order
$10^{-3}$ for $ m_{b'} > m_{t'}$, while $t'\to t\gamma$ is of
order a few $10^{-5}$. The $t'\to tH$, $tZ$ rates may be
measurable at the LHC.

\section{\boldmath FCNC $e^+e^-\to Q\bar{q}+\bar{Q}q$ Associated Production}

There has been several studies looking for collider signatures of
the FCNC top couplings, both at lepton colliders as well as at
hadron colliders. To the best of our knowledge, there is no
dedicated study of the associated production $e^+e^-\to
Q\bar{q}+\bar{Q}q$, with $Q$-$q$ being $t$-$c$, $t^\prime$-$t$,
$b^\prime$-$b$ or $b$-$s$. Because of the very large mass of the
heavy quark $Q$, $Q\bar{q}$ production at an $e^+e^-$ machine
would have a clear signature \cite{soni}. The top-charm production
at lepton colliders $e^+e^-$ and $\mu^+\mu^-$ has been studied in
two Higgs doublet models with and without Natural Flavor
Conservation, the so called 2HDM-II and 2HDM-III, which can lead
to measurable effects \cite{soni,sonimu,abdess}. It has been
pointed out that the tree level Higgs vertex $\phi \bar{t}c$ can
be better probed through $t$-channel $WW$ and/or $ZZ$ fusion at
high energy $e^+e^-$ collisions $e^+e^- \to t \bar{c}\nu_e
\bar{\nu}_e$ and $e^+e^- \to t \bar{c} e^+ e^-$ \cite{2hdm11,hou}.
An interesting feature of those reactions is that, being
$t$-channel, their cross sections grow with energy, unlike
$s$-channel reactions $e^+e^- \to t \bar{c}$, which are suppressed
at high energies. The cross sections of $e^+e^- \to t \bar{c}\nu_e
\bar{\nu}_e$ and $e^+e^- \to t \bar{c} e^+ e^-$ are found to be
one or two orders of magnitude higher than the cross sections of
$e^+e^- \to t \bar{c}$ \cite{2hdm11,hou}.

%%%%%%%%%%%%%%%%%%%%%%%%%%%%%
\begin{figure}[b!]
\smallskip\smallskip
%%%%%%%%%%%%%%%%%%%%%
\centerline{{ \epsfxsize5.5 in \epsffile{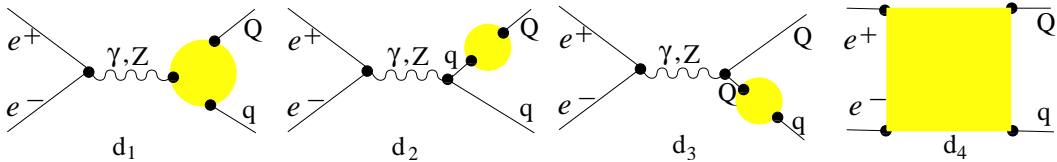}}}
\caption{Topological contributions to $e^+e^-\to Q\bar{q}$. }
\label{eee}
\end{figure}

With the above possible probes to Higgs FCNC couplings in the
backdrop, here we pursue the direct probe of FCNC in $e^+e^-$
collisions. We have three sets of diagrams for $e^+e^-\to
Q\bar{q}$ process: $e^+e^- \to \gamma^*\to Q\bar{q}$, $e^+e^- \to
Z^*\to Q\bar{q}$, and box diagrams, as depicted in Fig.~\ref{eee}.
Calculation of the full set of diagrams is done with the help of
FormCalc \cite{FA2}. We have checked both analytically and
numerically that the result is ultraviolet (UV) finite and
renormalization scale independent. We will present only
unpolarized cross sections. It is well known that the cross
sections for polarized initial states differ from the unpolarized
cross sections only by a normalization factor.

In the present study, we limit ourselves to $e^+e^-$ colliders.
The cross sections for $e^+e^- \to Q\bar{q}$, if sizeable, can
give information on the FCNC couplings $Q\to q\gamma $ and $Q \to
q Z$ as well as on $Z\to s\bar{b}$. We shall first consider
$e^+e^-\to b\bar s$ as it is the only one that can be probed in
principle at the high luminosity SuperB factories, and also at a
specialized $Z$ factory. In the second subsection we will turn to
the cases of $Q = t$, $b'$ and $t'$.

\subsection{\boldmath $e^+e^-\to b{\bar{s}}+{\bar{b}}s$}

As pointed in the Introduction, in SM the branching ratio for
$Z\to b{\bar{s}}+{\bar{b}}s$ is of the order $10^{-8}$. New
physics contributions, like 2HDM and SUSY, to rare $Z\to
b{\bar{s}}+{\bar{b}}s$ decay have been extensively studied
\cite{zsb1,zsb2} and shown to enhance ${\cal B}(Z\to
b{\bar{s}}+{\bar{b}}s)$. At LEP the sensitivity to rare Z decay is
about $10^{-5}$, while at future Linear Colliders operating at Z
mass (e.g. GigaZ option of the TESLA LC) will bring this
sensitivity up to the level of $10^{-8}$ \cite{ILC}. %TESLA}.
It is then legitimate to look for new physic in rare Z decays.
Ref.~\cite{zsb2} considered effect of 2HDM-III together with 3 and
4 fermions generations. In this section we will consider effect of
sequential fourth generation on rare Z decay using our
parameterization described before and taking into account
experimental constraints such as $B\to X_s \gamma$ and $B\to X_s
l^+l^-$.

%%%%%%%%%%%%%%%%%%%%%%%%%%%%%%
\begin{figure}[t!]
\smallskip\smallskip
\vskip-.1cm
%%%%%%%%%%%%%%%%%%%%
\centerline{{ \epsfxsize3.4 in \epsffile{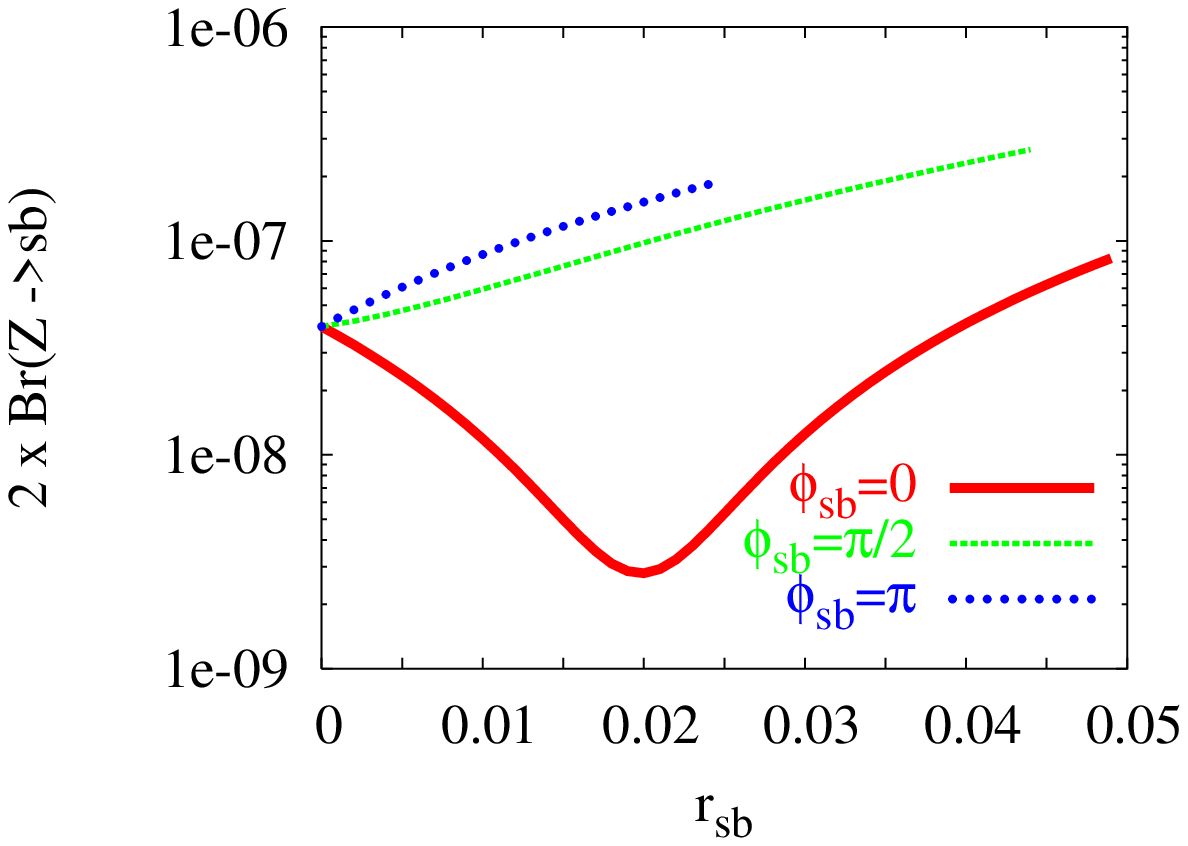}} \hskip-1.4cm
\epsfxsize3.4 in \epsffile{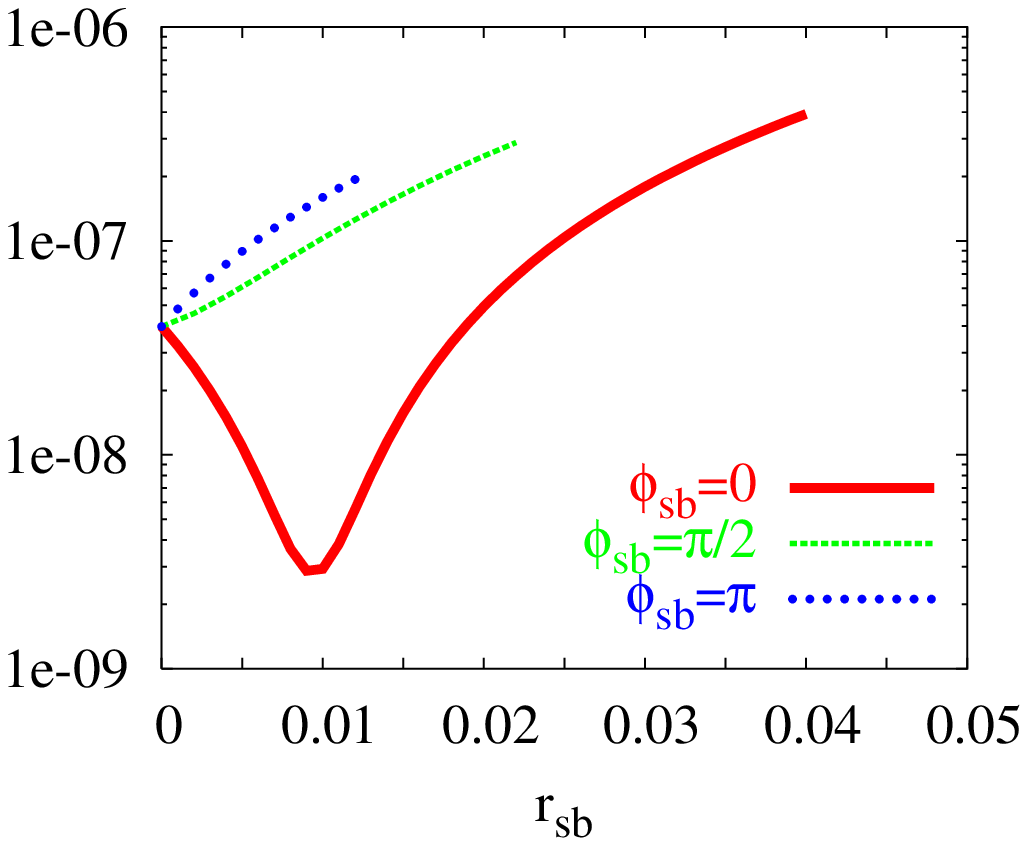}}
\smallskip\smallskip
\caption{Fourth generation contribution to ${\cal B}(Z\to
 b\bar{s}+\bar{b}s)$ vs $r_{sb}$ for $m_{t^\prime}=300$
 GeV (left) and $400$ GeV (right) for three
values of $\phi_{sb}=0, \pi/2, \pi$. The lines terminate when
$B\to X_sl^+l^-$ exceeds 2$\sigma$ range from experimental value
of $(6.1_{-1.8}^{+2.0}) \times 10^{-6}$.}
\label{plot9}
\end{figure}

The amplitude of such $b\to s$ transition is of the form given by
Eq.~(\ref{asb}). Like the case of $b\to s \gamma$, the $CP$ phase
$\phi_{sb}$ could play a crucial role. In Fig.~\ref{plot9} we
illustrate fourth generation contribution to the branching ratio
of $Z\to b\bar{s}$ as function of $r_{sb}$ for $m_{t^\prime}=300$
GeV (left) and $400$ GeV (right), and for three values of
$\phi_{sb}=0, \frac{\pi}{2}, \pi $. We allow $B\to X_sl^+l^-$ to
be in the $2\sigma$ range of the experimental value
$(6.1_{-1.8}^{+2.0}) \times 10^{-6}$. Data points which do not
satisfy this constraint are not plotted. One sees that ${\cal
B}(Z\to b\bar{s})$ can reach $4\times 10^{-7}$ for $r_{sb} \approx
0.04$ and $m_{t^\prime}=400$ GeV. The observed dip in the plots
correspond to destructive interference between $t$ and $t^\prime$
contributions. This dip appears only for $\phi_{sb}=0$, while for
$CP$ phase $\phi_{sb}\ga \pi/2$ the $t'$ contribution interferes
constructively with the top which leads to a small enhancement of
the rate (see Eq.~(\ref{asb})).

%%%%%%%%%%%%%%%%%%%%%%%%%%%%%%
\begin{figure}[t!]
\smallskip\smallskip
%%%%%%%%%%%%%%%%%%%%%
\centerline{{ \epsfxsize3.4 in \epsffile{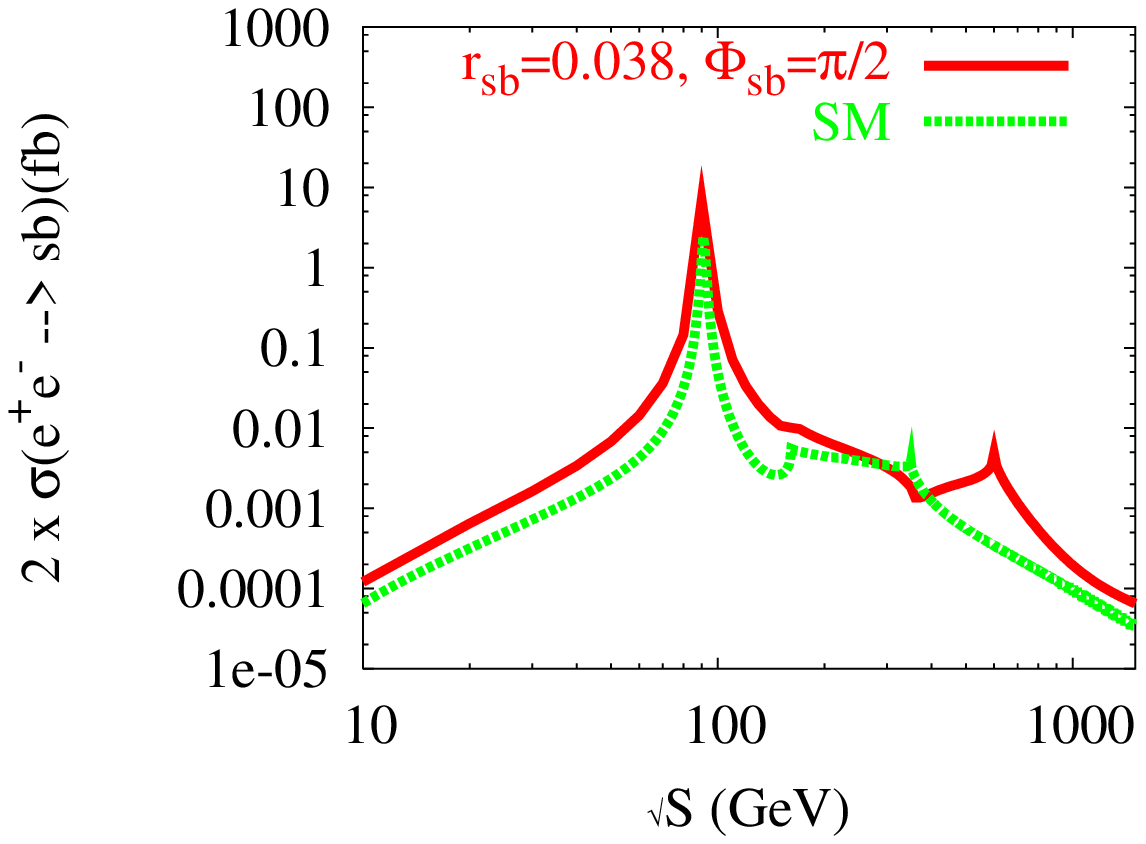}}  \hskip-1.6cm
\epsfxsize3.4 in \epsffile{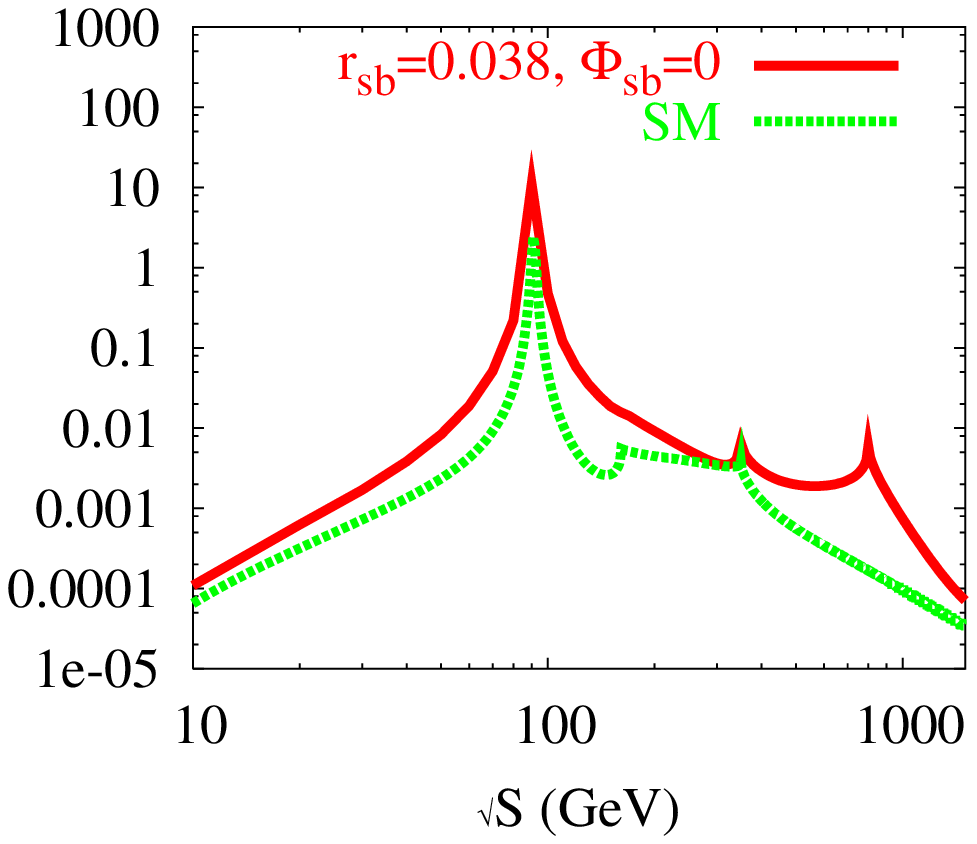} }
\smallskip\smallskip
\caption{Fourth generation contribution to $e^+e^-\to
  s\bar{b}+\bar{s}b$ as a function of
$\sqrt{s}$  for $r_{sb}=0.038$, $m_{t^\prime}=300$ GeV (left),
$400$ GeV (right).}
\label{plot10}
\end{figure}

Off the $Z$ peak, $e^+e^-\to \bar{b}s$ can still be probed at the
future ILC. The cross section can be enhanced by about one order
of magnitude with respect to SM cross section, depending on
thresholds and the relative phase $\phi_{sb}$, as illustrated in
Fig.~\ref{plot10}. The values of $r_{sb}$ and $\phi_{sb}$ have been
fixed such that $B\to X_sl^+l^-$ is satisfied. In fact, we choose
$r_{sb}$ and $\phi_{sb}$ that saturate ${\cal B}(B\to X_sl^+l^-)$.
Near the $Z$ pole $\sqrt{s}\approx M_Z$, the cross section can be
larger than 10 fb and decreases when we increase the CM energy.
One can reach $\approx 0.01$ fb for $m_{t^\prime}=400$ GeV and
$r_{sb}=0.038$. There are several threshold effects which manifest
themselves as small kinks in Fig.~\ref{plot10}, corresponding to
$WW$, $tt$ and $t^\prime t^\prime$ threshold production.

Motivated by the high luminosity accumulated by the $B$ factories,
e.g. about 500 fb$^{-1}$ by Belle experiment at present, as well
as the possibility for a future SuperB factory upgrade, we study
the associated production of bottom-strange at center of mass
energy $\sqrt{s} \simeq 10.6$ GeV both in SM and in SM with fourth
generation. This is illustrated also in Fig.~\ref{plot10}, which
extends down to $\sqrt{s}=10.6$ GeV. In SM one has about $10^{-4}$
fb which leads to negligible number of events at present $B$
factories. With fourth generation contribution, the cross section
for $e^+e^-\to {\bar{b}}s$ at $\sqrt{s}=10.6$ GeV can be enhanced
only mildly, largely due to $B\to X_sl^+l^-$ constraint, and
remains of the order $10^{-4}$. But with several order of
magnitude increase in luminosity at the SuperB factories,
$e^+e^-\to {\bar{b}}s$ may become interesting.

\subsection{\boldmath $e^+e^- \to t\bar c, b'\bar b, t'\bar t$}

\begin{figure}[t!]
\smallskip\smallskip
%%%%%%%%%%%%%%%%%%%%%
\centerline{{ \epsfxsize3.4 in \epsffile{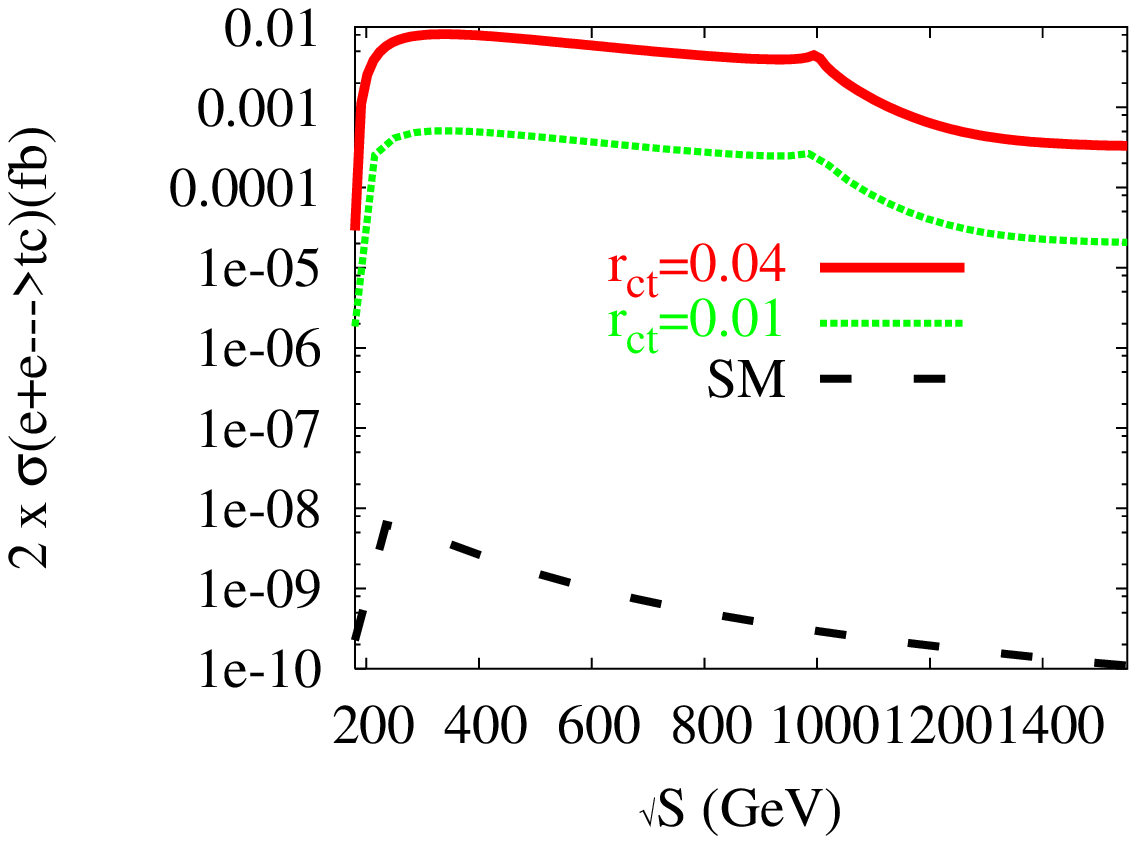}}  \hskip-1.6cm
\epsfxsize3.4 in \epsffile{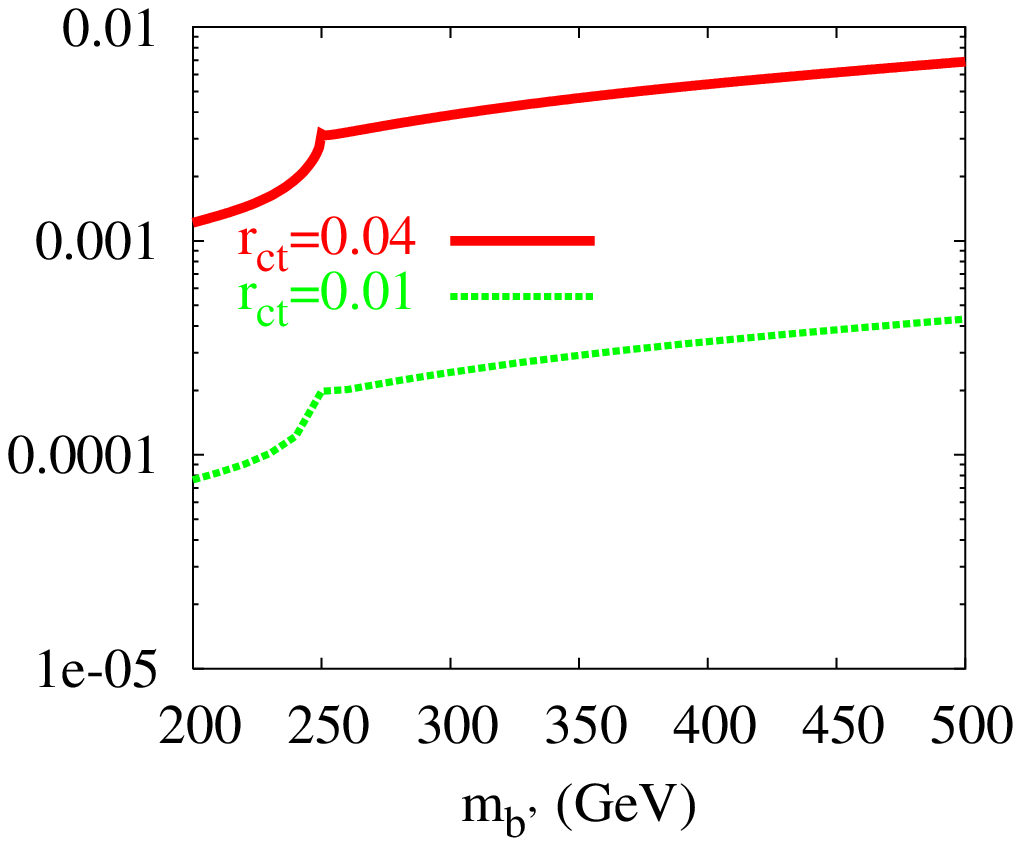} }
\smallskip\smallskip
\caption{
 Fourth generation contribution to $e^+e^-\to t\bar{c}+\bar{t}c$
 as a function of $\sqrt{s}$ (left) for $m_{b'} = 500$ GeV, and
 as a function of $m_{b^\prime}$ at $\sqrt{s}=500$ GeV (right)
 and 2 values of $r_{b^\prime}$.
 The SM cross section for $e^+e^-\to t\bar{c}+\bar{t}c$ is
 also shown in left plot. }
\label{plot11}
\end{figure}

Due to severe GIM cancellations between bottom, strange and down
quarks as their masses are close to degenerate on the top scale,
the SM cross section for $e^+e^- \to \bar{t}c$ is very suppressed.
As one can see from Fig.~\ref{plot11}(a), the $e^+e^- \to \bar{t}c$
cross section in SM is more than five orders of magnitude lower
than the corresponding $e^+e^- \to \bar{b}s$ cross section in SM
given in Fig.~\ref{plot10} at the same energy. The cross section is
of order $10^{-9}$ fb, in agreement with \cite{9901369}. For
$e^+e^- \to \bar{b}s$ the internal fermions in SM are top, charm
and up quarks. Because of the large top mass which is well split
from the other internal quarks, the cross section for $e^+e^-\to
\bar{b}s$ is in the range $10^{-4}$--$10^{-2}$ fb for $\sqrt{s}\in
[200,1000]$ GeV as already discussed, and in agreement with
\cite{9902474}.

The fourth generation $b'$ quark can clearly affect the $e^+e^-\to
\bar{t}c$ cross section significantly, as seen already in our
discussion of $t\to cX$ decays. For sake of future ILC studies, we
illustrate the cross section vs $\sqrt{s}$ in Fig.~\ref{plot11}(a)
our numerical results for heavy $m_{b'}=500$ GeV and several
values of $r_{ct}$. We also show the sensitivity to $m_{b'}$ mass
in Fig.~\ref{plot11}(b).
We see that the cross section for $e^+e^-\to \bar{t}c$ can get
enhanced by six orders of magnitude with respect to the SM values.
The cross section turns on sharply above threshold, becoming
sizeable above $\sqrt{s}=200$ GeV, and can reach values of $\sim
0.01$ fb for large $r_{ct}=0.04$. One would still need a high
luminosity of ${\cal L}\ga 500$ fb$^{-1}$ or more to get at best a
few events. Above $\sqrt{s}\ga 1000$ GeV, the cross section
decreases with increasing energy, reaching a value of $\approx
10^{-4}$ fb at $\sqrt{s}\ga 1.5$ TeV. One can see a kink around
$\sqrt{s}=1000$ GeV in Fig.~\ref{plot11}(a), which corresponds to
threshold production of $b^\prime$ pair.
In Fig.~\ref{plot11}(b) we illustrate the dependence of $e^+e^-\to
\bar tc$ cross section on $b^\prime$ mass for the more modest
energy of $\sqrt{s}=500$ GeV. At $m_{b^\prime}=250$ GeV, one can
see a kink which corresponds to the threshold production of
$b^\prime$ pair. For $r_{ct}=0.04$ the cross section can be
enhanced by one order of magnitude when varying $m_{b^\prime}$
from 200 to 500 GeV.

%%%%%%%%%%%%%%%%%%%%%%%%%%%%%%

By the same reason that $b' \to b$ and $t'\to t$ transitions have
larger rates than $t\to c$ transitions, one expects $e^+e^-\to b'
\bar b$ and $t'\bar t$ to have larger cross sections. If the
fourth generation exists, it would be copiously produced at the
LHC, and discovery is not a problem. For the future ILC, the
simplest way to produce fourth generation $Q$ is through $Q$ pair
production $e^+e^-\to \gamma^*,Z^*\to Q\bar{Q}$ if enough center
of mass energy is available, i.e. $\sqrt{s}\ga 2 m_Q$. But the
production of fourth generation $Q$ in association with a lighter
quark $q$, $e^+e^- \to Q\bar q$, would be kinematically better
than $e^+e^-\to Q\bar{Q}$ pair production. It offers the
possibility of searching for $m_Q$ up to $\sqrt{s}-m_q$, in
contrast to pair production which only probes up to $m_Q\la
\sqrt{s}/2$.

%%%%%%%%%%%%%%%%%%%%%%
\begin{figure}[t!]
\smallskip\smallskip
%%%%%%%%%%%%%%%%%%%%%
\centerline{{
\epsfxsize3.4 in
\epsffile{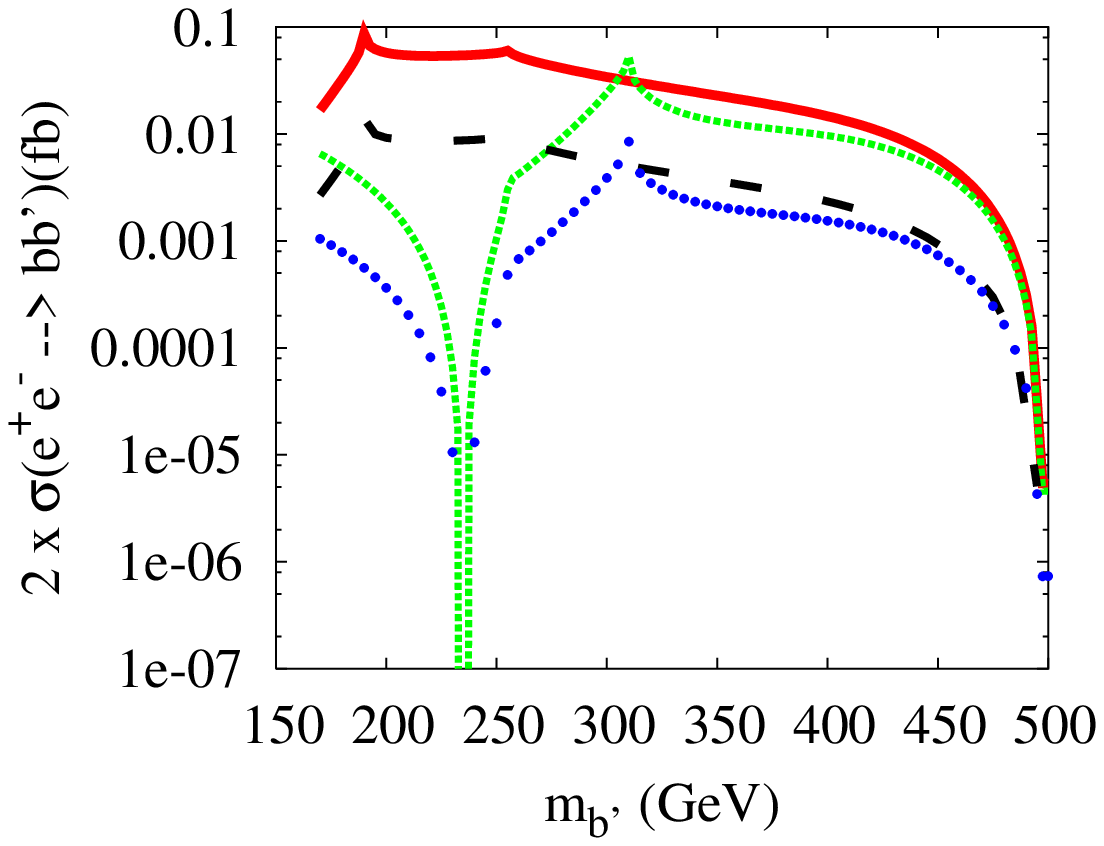}}  \hskip-1.6cm
\epsfxsize3.4 in
\epsffile{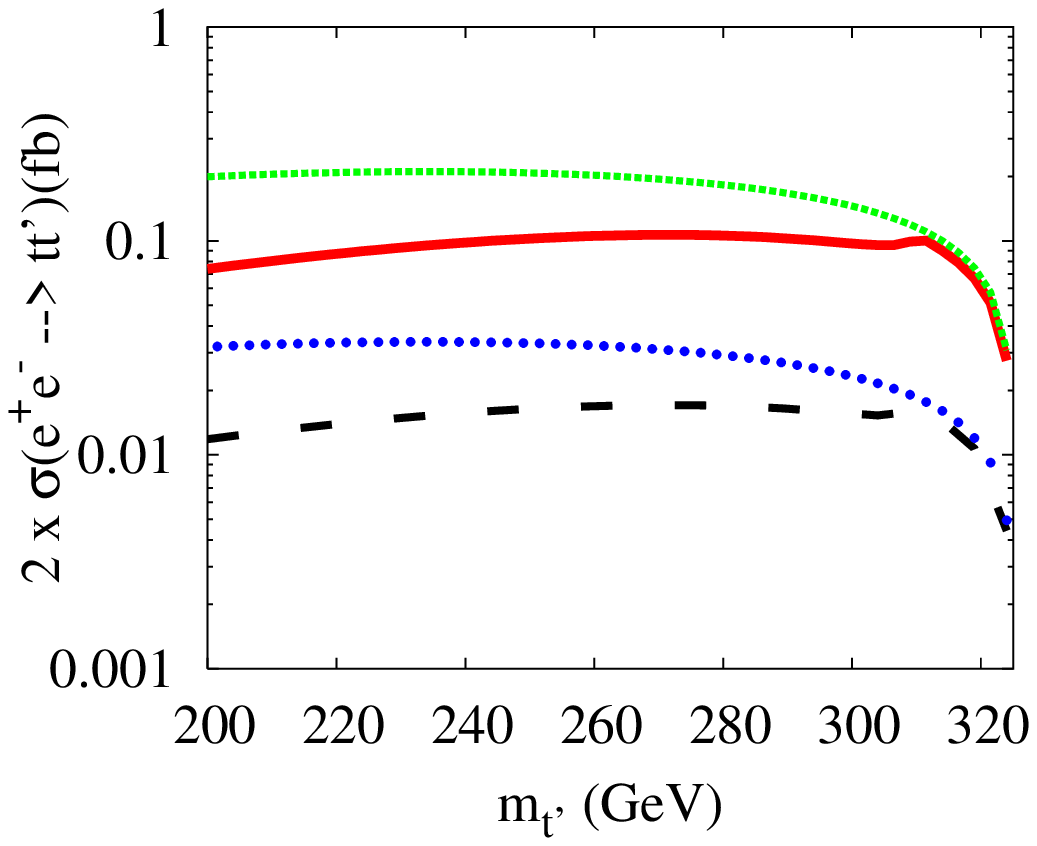} }
\smallskip\smallskip
\caption{Fourth generation contribution to $e^+e^-\to
  b\bar{b'}$ as a function $m_{b'}$ (left), and $e^+e^-\to
  t\bar{t'}$) as a function of
  $m_{t'}$ (right) at $\sqrt{s}=500$ GeV.
$r_{bb'}=0.25$ for solid and small dash plots and
$r_{bb'}=0.1$ for long dash and dots plots.
$m_{t'}=m_{b'}+\Delta $ GeV with $\Delta=60$ GeV for solid and long
  dash and -60 GeV for the other plots}
\label{plot12}
\end{figure}
%%%%%%%%%%%%%%%%%%%%%%%%%%%%%%%
%%%%%%%%%%%%%%%%%%%%%%%%%%%%%%
\begin{figure}[t!]
\smallskip\smallskip
%%%%%%%%%%%%%%%%%%%%%
\centerline{{
\epsfxsize3.4 in
\epsffile{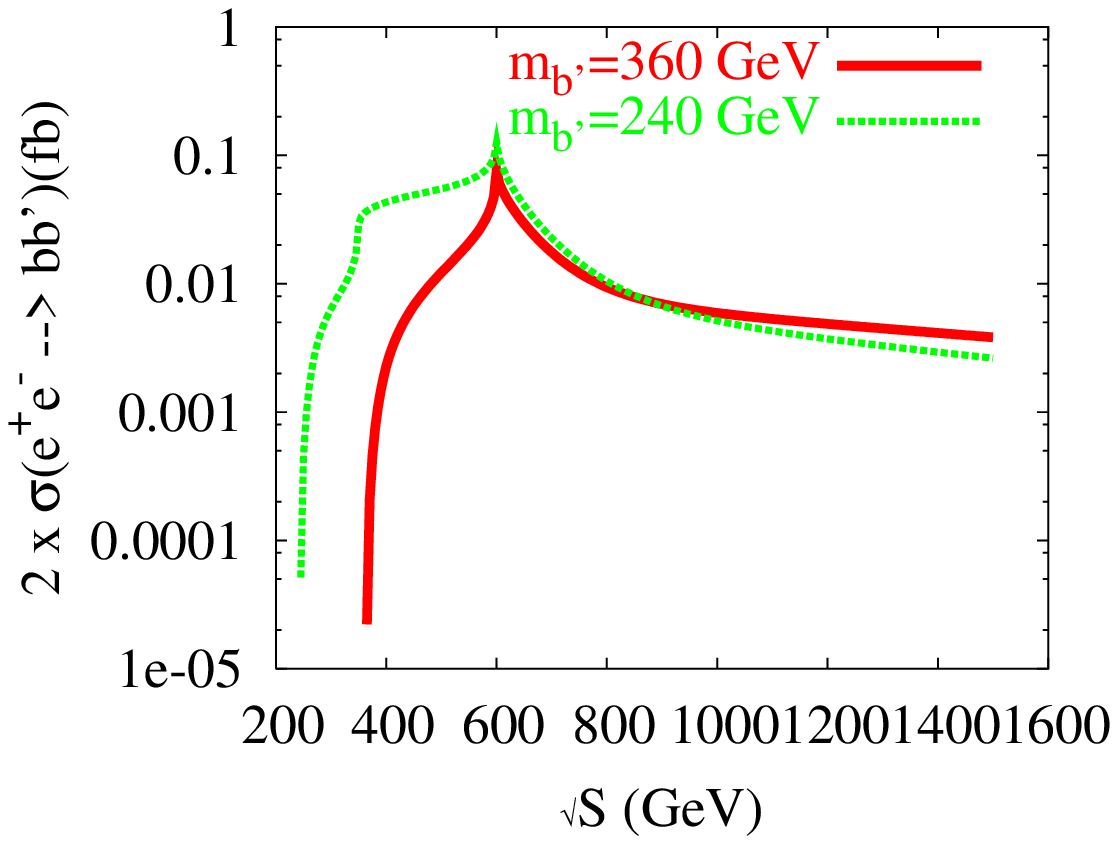}}  \hskip-1.6cm
\epsfxsize3.4 in
\epsffile{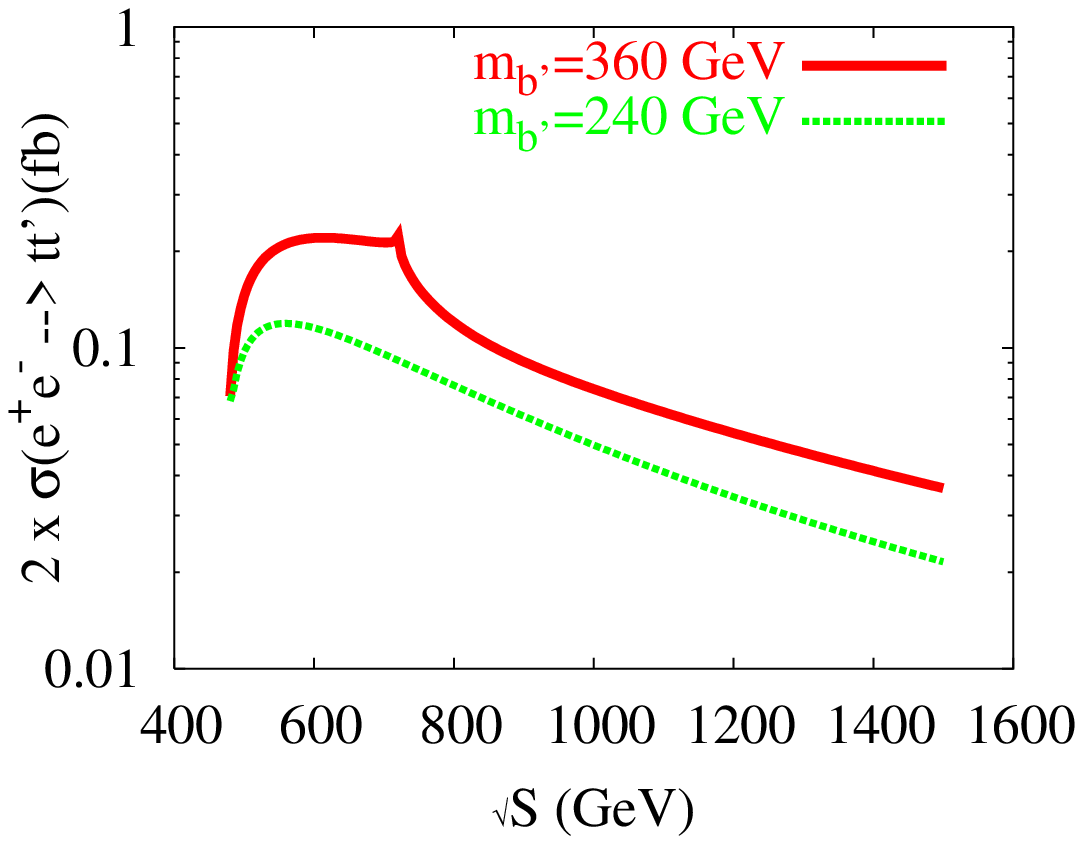} }
\smallskip\smallskip
\caption{$\sigma(e^+e^-\to
  b\bar{b'}) $  (left) and $\sigma(e^+e^-\to
  t\bar{t'}) $ (right) as a function $\sqrt{s}$
for $r_{bb'}=r_{tt'}=0.25$, $m_{t'}=300$ GeV and
$m_{b'}$ as indicated}
\label{plot13}
\end{figure}

In Fig.~\ref{plot12}, we present cross sections for both $e^+e^-\to
b\bar{b'}$ (left) and $t\bar{t'}$ (right) at $\sqrt{s}=500$ GeV,
as functions of $m_{b^\prime}$ and $m_{t^\prime}$, respectively.
Obviously, the largest $r_{bb'}$ (or $r_{tt'}$) gives the largest
cross section. We have illustrated both cases with
$m_{t'}=m_{b'}\pm 60$ GeV.
In the case where $m_{t'}=m_{b'}+ 60$ GeV $>m_{b'}$,
this explains the kink at $m_{b'} = 190$
GeV in Fig.~\ref{plot12}(a), which corresponds to the opening of the
$t'\bar t'$ threshold, and a slight kink at $m_{b'} = 255$ GeV,
which corresponds to the $b' \to tW$ cut.
While in the case $m_{t'}=m_{b'}- 60$ GeV $<m_{b'}$,
its clear that for $m_{b'}=235$ GeV which correspond to
$m_{t'}=175$ GeV $=m_t$ there is a GIM cancellation, while the kink around
$m_{b'} \approx$ 310 GeV corresponds to the opening of the
$t'\bar t'$ threshold.
Above $m_{b'}\approx 250$ GeV, $b'\bar b'$ pair production no
longer occurs, but $e^+e^-\to b\bar{b'}$ can still be probed,
with cross section of order a few 0.01 fb. Analogously,
as seen in Fig.~\ref{plot12}(b), the cross
section for $e^+e^-\to t\bar{t'}$ remains more or less constant
around 0.2 fb, even when $t'\bar t'$ pair production is forbidden.
Of course, the possibilities are richer as $m_{b'}$ and $m_{t'}$
are unknown, but in both $e^+e^-\to b\bar{b'}$ and $e^+e^-\to
t\bar{t'}$ cases, it is clear that for large $r_{bb'} \sim 0.2$,
the cross sections can be larger than $0.01$ fb, reaching 0.1 fb,
which could lead to a few ten events for the high
luminosity option ${\cal L}\ga 500\ {\rm fb}^{-1}$.\\
For completeness, we show in Fig.~\ref{plot13}
cross sections for both $e^+e^-\to
b\bar{b'}$ (left) and $t\bar{t'}$ (right) for $m_{t'}=300$ GeV,
$m_{b'}=240, 360$ GeV and large $r_{bb'}=r_{tt'}=0.25$
as functions of center of mass energy, respectively.
It is clear from Fig.~\ref{plot13} that the associate production
$e^+e^-\to b\bar{b'}$ or $e^+e^-\to
t\bar{t'}$ can lead to a few ten of event before one can accumulate
enough energy to produce a pair of $b'$ or $t'$.
Thus, one can not only study FCNC $b'\to bZ$ and $t'\to tX$ decays
at the ILC, one could also probe FCNC induced associated $b'\bar
b$ and $t'\bar t$ production.

%%%%%%%%%%%%%%%%%%%%%%%%%%%%%%%%%%%%%%%%
\section{Discussion}

At the Tevatron or the LHC, fourth generation $b'$ and $t'$ quarks
can be pair produced through $gg$ fusion and $q\bar{q}$
annihilation, with the same sizeable QCD cross section for a given
mass. For example, at Tevatron, for $\sqrt{s}=2$ TeV and a heavy
quark $m_Q=240$ GeV, one can have a non-negligible cross section
of the order $\approx 1.2$ pb. At the LHC the cross section can
increase by about 2 orders of magnitude, with much higher
luminosity. Consequently, discovering the fourth generation is not
a problem for hadronic machines, and we are on the verge of
finally discovering the fourth generation, or ruling it out
conclusively. The search strategy would depend on decay pattern,
and once discovered, the ILC may be needed for precise
measurements of all sizable decay modes.

We are concerned with heavy fourth generation $m_{b^\prime},
m_{t^\prime}\ga 200$ GeV. The allowed tree level decays of the
$b^\prime$ are $b^\prime \to cW$ (ignoring $uW$), $tW^{(*)}$, and
$t^\prime W^*$ if kinematically possible, and of course our main
interest of FCNC decays $b^\prime \to bZ, bH, bg, b\gamma$, and
the suppressed $b^\prime \to sZ, sH, sg, s\gamma$ (see Appendix).
Similarly, the allowed $t^\prime$ tree level decays are $t^\prime
\to sW, bW$ (dropping again $dW$), $b^\prime W^*$ and the
loop-induced FCNC decays $t^\prime \to tZ, tH, tg, t\gamma$ and
$t^\prime \to cZ, cH, cg, c\gamma$.
The relative weights of these decays have been surveyed in
Section~4, and the prognosis is that FCNC decays can be measured
at the LHC, once the fourth generation is discovered. Here we
offer some discussion on various special situations.

\subsection{Comment on Tevatron Run-II}

For $m_t + M_W> m_{b^\prime} > M_Z+ m_b$ and if $V_{cb'}$ is
suppressed, the decay $b^\prime \to b Z$~\cite{hou1} is expected
to dominate over the other FCNC decay processes, except for
$b^\prime \to b H$~\cite{HR,sonibp} if $m_{b^\prime} > M_H+ m_b$
also. The CDF Collaboration \cite{cdf4} gave an upper limit on the
product $\sigma(p\bar{p}\to b^\prime\bar{b^\prime})
 \times [{\cal B}(b^\prime\to bZ)]^2$
as a function of $m_{b^\prime}$,
which excludes at 95\% CL the range 100 GeV $ < m_{b^\prime} <$ 199 GeV
if ${\cal B}(b^\prime\rightarrow bZ) = 100\%$.
For ${\cal B}(b^\prime \to b H) \neq 0$,
so long that ${\cal B}(b^\prime \to b Z)$ does not vanish,
the CDF bound still largely applies
since hadronic final states of $b^\prime \to b Z$ and $bH$
are rather similar, and in fact the $bH$ mode
has better $b$-tagging efficiency.

What CDF apparently did not pursue in any detail is the $b^\prime
\to cW$ possibility. Nor has the complicated case of $b' \to tW^*$
been much discussed. Clearly the $b$-tagging efficiency for $cW$
mode would be much worse than $bZ$ or $bH$. Since $b$-tagging is
an important part of the CDF $b^\prime \to b Z$ search strategy,
the CDF search may be evaded if ${\cal B}(b^\prime \to cW)$ is
sizable. As we have demonstrated in Fig.~2, $b'\to cW$ can be the
dominant decay mode for $m_{b'}=240$ GeV for all range of
$r_{bb'}$, if $V_{cb'}$ could be as sizable as $V_{cb} \simeq
0.04$. However, in case $V_{cb'}$ is rather suppressed, FCNC
decays like $b'\to bZ, bH$ can compete with $b'\to cW$, $tW^*$.
Hence, more generally one should make a combined search for $b'
\to cW, bZ$ and $bH$ \cite{AH}, even $tW^*$. On the other hand,
$t'$ could turn out to be lighter than $b'$. In this case one
expects top-like decay pattern, and one can look for $t^\prime \to
bW$, but $t'\to sW$ could also be sizable, again diluting the
effectiveness of $b$-tagging.

We stress that, for either the case of sizable $b'\to cW$ or
$t'\to sW$, the possible loss of $b$-tagging efficiency should be
kept in mind in heavy quark search. Although one may face more
background because of this, the possibility of dominant or
prominent FCNC $b'\to bZ$, $bH$ should continue to be pursued in
the closing years of Tevatron Run-II.

%%%%%%%%%%%%%%%%%%%%%%%%%%%%
\subsection{Large or Small \boldmath $V_{cb'}$ ($V_{t's}$)?}

The CKM matrix becomes $4\times 4$ in case of four generations.
Since the elements $V_{ud}$, $V_{us}$, $V_{cd},$ $V_{cb},$
$V_{cs}$, $V_{ub}$, in that order, are suitably well measured, one
can continue to use the usual three generation parameterization.
The additional 3 mixing angles and 2 $CP$ phases, following
Ref.~\cite{HSS}, are placed in $|V_{t'b}|$, $|V_{t's}|$,
$|V_{t'd}|$, and $\arg (-V_{t's})$, $\arg (-V_{t'd})$. Without
much loss of generality for future heavy quark search and studies,
we have dropped $V_{t'd}$, as well as $V_{ub'}$ \cite{vubp}. It is
nontrivial that we already have a bound on $|V_{t'b}|$, as given
in Eq. (\ref{Zbb}). Thus, the main unknown for our purpose is the
angle and phase in $V_{t's}$ (which is closely related to
$V_{cb'}$). This element could impact on $b\to s$ transitions if
sizable.

\subsection*{Large \boldmath $V_{t's}$ Scenario }

There are two recent hints of New Physics in $b\to s$ transitions.
One is the difference, called $\Delta {\cal S}$, between the
time-dependent $CP$ violation measured in penguin dominant $b\to
s\bar ss$ modes such as $B\to \phi K_S$, and tree dominant $b\to
c\bar cs$ modes such as $B\to J/\psi K_S$. Three generation SM
predicts  $\Delta {\cal S} \cong 0$ \cite{Beneke}, but
measurements at B factories persistently give $\Delta {\cal S} <
0$~\cite{HFAG} in many modes, though they are not yet
statistically significant. The other indication is the difference
in direct $CP$ violation measured in $B \to K^+\pi^-$ vs
$K^+\pi^-$ modes~\cite{HFAG}, ${\cal A}_{K^+\pi^0} - {\cal
A}_{K^+\pi^-} \neq 0$. Although this could be due to hadronic
effects such as enhancement of so-called ``color-suppressed"
amplitude, the other possibility could be New Physics in the
electroweak penguin amplitude. It has been shown that the fourth
generation contribution to EW penguin with large
$V_{t's}^*V_{t'b}$ and near maximal $CP$ phase~\cite{HNS} could
explain the effect. It is consistent with $B$, $K$ and $D$ data,
and predicts enhanced $K_L\to \pi^0\nu\bar\nu$
decay~\cite{HNS_KL}. Assuming $m_{t'}=300$ GeV, it turns out that
large $V_{t's}^*V_{t'b}\approx 0.025$ with large phase is allowed,
while $|V_{t's}|$ is only slightly smaller than $|V_{t'b}|$. It
further gives rise to the downward trend of $\Delta {\cal S} < 0$
observed at B factories, though not quite sufficient in
strength~\cite{HNRS}. Thus, it could in principle account for both
hints of New Physics. As we have demonstrated, after taking into
account $b\to sl^+l^-$ constraint, $Z\to sb$ is only slightly
enhanced. But Ref.~\cite{HNS} predicts large and negative $CP$
violation in $B_s$ mixing, which can be probed at the Tevatron
Run-II, and can be definitely measured by LHCb at the LHC shortly
after turn-on.

For our purpose of heavy quarks decays, taking $V_{t'b} \simeq
-0.22$, Ref.~\cite{HNS_KL} finds
\begin{eqnarray}
 && V_{cb'} \simeq 0.12 \, e^{i66^\circ}, \ \ V_{t's}^*
\simeq -0.11 \, e^{i70^\circ}, \ \
    V_{tb'} \simeq 0.22 \, e^{-i1^\circ}, \nonumber \\
 && V_{t's}^*V_{t'b} \equiv r_{sb}\, e^{i\phi_{sb}}
  \simeq 0.025\, e^{i70^\circ}, \ \
 V_{cb'}V_{tb'}^* \equiv r_{ct}\, e^{i\phi_{ct}}
  \simeq 0.025 e^{i67^\circ}, \nonumber \\
 && V_{t'b}^*V_{t'b'} \equiv r_{bb'} \simeq -0.21, \ \
 V_{tb'}V_{t'b'}^* \equiv r_{tt'}
  \simeq 0.21 \, e^{-i1^\circ}.
 \label{HNS}
\end{eqnarray}
Thus, $V_{cb'}$ and $V_{t's}$ are even larger than the 0.04 value
used in Figs. \ref{plot4} and \ref{plot6}. If the scenario is
realized, $b'\to cW$ would predominate for $m_{b'} \la m_{t'}$,
and even for $m_{b'} > m_{t'}$, $b'\to cW$ would be comparable to
$b'\to tW$. The relative FCNC rates would be slightly reduced, but
still measurable. The suppressed $b'\to sX$ decays discussed in
Appendix may become interesting. Since the $\Delta {\cal S}$ and
${\cal A}_{K^+\pi^0} - {\cal A}_{K^+\pi^-} \neq 0$ problems may
soften, we have only presented the $V_{cb'} \approx V_{cb} \approx
0.04$ case in Figs. \ref{plot4} and \ref{plot6}.

\subsection*{Very Suppressed \boldmath $|V_{cb'}|\approx
|V_{t^\prime s}|$ Scenario}

The measured three generation quark mixing elements exhibit an
intriguing pattern of $|V_{ub}|^2 \ll |V_{cb}|^2 \ll |V_{us}|^2
\ll 1$. In the four generation scenario of Ref.~\cite{HNS,HNS_KL},
this pattern is violated by the strength of $|V_{cb'}|^2$ in
Eq.~(\ref{HNS}) being an order of magnitude larger than
$|V_{cb}|^2$. This is in itself not a problem, since the CKM
mixing elements are parameters of the SM, and are {\it a priori}
unknown. The $b\to s$ transitions may well hold surprises for us.

It may also happen that the hints for New Physics in $b\to s$
transitions eventually evaporate. For that purpose, we have shown
the other end of very suppressed $|V_{cb'}|\approx |V_{t^\prime
s}| \approx 10^{-3}$.
Part of the reason for choosing such a small value is because, for
$m_{b'}\la 200$ GeV which is well within reach of the Tevatron, it
has been extensively discussed in Refs.~\cite{AH,HR}. It has been
shown that, if we take the ratio $|V_{cb^\prime }/(V_{t^\prime
b^\prime} V_{t^\prime b}^*)|$ to be of the order $10^{-3}$, then
the tree level $b^\prime \to cW$ decay and the loop level
$b^\prime \to bZ, bH$ decays could be comparable, and the CDF
bound can be relaxed.
As seen in Figs. \ref{plot5} and \ref{plot7}, for even smaller
$V_{cb'}$, the $b'\to cW$ and $t'\to sW$ decays become rare
decays. This is analogous to tree dominant $b\to u$ decays such as
$B\to \pi^+\pi^-$ being weaker than loop induced $b\to s$ decays
such as $B \to K^+\pi^-$.

For very small $V_{cb'}$ (with $V_{ub'}$ already assumed small),
one effectively has the $2\times 2$ structure between the two $(t,
b)_L$ and $(t', b')_L$ doublets, with the mixing element $V_{t'b}
\cong -V_{tb'}$ controlling both the tree level and loop induced
decays. The phenomenology has already been considered in
Section~4. In general one expects $b'\to tW$ and $t'\to bW$ to be
dominant, with FCNC rates at $10^{-4}$ to $10^{-2}$ level. The
exception is when $m_{b'} \la m_t + M_W$, so $b'\to tW^*$ gets
kinematically suppressed, as illustrated in Fig.~\ref{plot5}(a).
This ``FCNC dominance" scenario is precisely the domain that is
relevant for Tevatron Run-II stressed in previous subsection.
Whether Tevatron explores this domain further or not, it would be
fully covered by the LHC.

The discovery of FCNC dominance for a {\it heavy} quark would be
truly amusing.

\section{Conclusions}

The study of flavor changing neutral couplings involving $s$ and
$b$ quarks has yielded a most fruitful program in the past 40
years, and is still going strong. With the turning on of the LHC
approaching, we will soon enter an era of studying FCNC involving
genuine heavy quarks, starting with the top. The (three
generation) SM predictions for $t\to cX$ where $X=Z$, $H$, $g$ and
$\gamma$ are orders of magnitude below sensitivity. This implies
an enormous range of probing for beyond SM(3) effects.

With the top quark as the single heavy quark at the weak scale, it
may be useful to contemplate additional heavy quarks such as a
sequential fourth generation. Unfortunately, we find that virtual
effects from $b'$ still cannot enhance $t\to cX$ to within
sensitivity at LHC, once $b\to s$ constraints are imposed. In the
case of $Z\to \bar sb$, again because of $b\to s$ constraints, we
find ${\cal B}(Z\to sb)$ can only reach $10^{-7}$. This at best
can be probed in the distance future at a specialized GigaZ.

The decay of fourth generation $b'$ and $t'$ quarks themselves are
far more promising. We have studied both the CC decays as well as
the FCNC decay modes.
We have shown that there is a rather broad range of possibilities
for $b'$ decay, depending on $m_{b'}$, $V_{cb'}$ and $V_{tb'}$.
For $m_{b'} \la 240$ GeV and very small $V_{cb'}$, FCNC dominance
is possible. In general, FCNC $b'\to bZ$, $bH$ decays could
compete with $b'\to cW$ and $tW^*$, and should be of interest at
the Tevatron Run-II. For sizable $V_{cb'}$ values, the $b'\to cW$
mode would dominate, which would greatly affect the effectiveness
of $b$-tagging for heavy quark search. For $m_{b'}
> m_t + M_W$, the dominance of $b'\to tW$ implies $b'\bar b' \to
t\bar tW^+W^- \to b\bar bW^+W^+W^-W^-$, or 4$W$s plus 2 $b$-jets,
which should be of interest at LHC. Except for the case of small
$V_{cb'}$ and light $m_{b'}$, the FCNCs are typically at
$10^{-4}$--$10^{-2}$ order hence always within reach at the LHC,
even though the signals may vary in richness and complexity.

The $t'$ case is simpler. Basically $t'\to bW$ dominates, so it
acts as a heavy top. In principle, $t'\to sW$ could cut in and
dilute $b$-tagging effectiveness, but that would require a rather
large $V_{t's}$ compared to $V_{t'b}$. The good news is that FCNC
rates are again in the accessible range at the LHC, with $t'\to
tH$ possibly reaching up to $10^{-3}$.

We have also studied the direct production through FCNC,
$e^+e^-\to q\bar{Q}$, at the future ILC. Unfortunately, these are
not very prominent. The $e^+e^-\to \bar sb$ and $\bar ct$ are
below sensitivity, while $e^+e^-\to \bar bb'$, $\bar tt'$ would
yield not more than a few ten events. Clearly, these numbers would
be better clarified once the fourth generation is discovered at
the LHC.

We conclude that Tevatron Run-II should be able to probe the light
${b'}$ (and $t'$ as well) case, where FCNC could be most
prominent. The LHC, however, should be able to establish the
fourth generation beyond any doubt, if it exists. Furthermore, the
LHC has the capability to measure all the $b'\to bZ$, $bH$, $bg$,
$b\gamma$ as well as $t'\to tZ$, $tH$, $tg$, $t\gamma$ decays and
offer a wealth of information. These modes could then be studied
in further detail at the ILC. Alternatively, the fourth generation
could {\it finally} be put to rest by the LHC.

\vspace{1cm} \noindent {\Large \bf Acknowledgements}\\
AA is supported by the Physics Division of National Center for
Theoretical Sciences under a grant from the National Science
Council of Taiwan. The work of WSH is supported in part by
NSC-94-2112-M-002-035. We are grateful to A. Soddu for bringing
Ref.~\cite{ADLO2005} to our attention.

\vskip1cm

\appendix

\section{Suppressed FCNC \boldmath $b'$ and $t'$ Decays}

The branching ratios for both the FCNC $b^\prime\to bX$ and
$t^\prime\to tX$ transitions are considerably larger than the
corresponding $t\to cX$ transitions, and should be observable at
the LHC. This comes about because they involve $r_{bb'} =
V_{t'b}^*V_{t'b'}$ and $r_{tt'} = V_{tb'}V_{t'b'}^*$, which are
larger than $r_{ct} = V_{cb'}V_{tb'}^*$ for $t\to cX$ transitions.
The leading tree level decays are controlled by $V_{(c)tb'}$,
$V_{t'b}$ and $V_{tb}$, respectively, which further suppress the
FCNC $t\to cX$ branching ratios. That is, the tree level $t\to bW$
decay is unsuppressed, while loop induced FCNC $t\to cX$ decays
are CKM suppressed. Having the decaying particles $b^\prime$ and
$t^\prime$ heavier than the top quark also helps the respective
$b^\prime$ and $t^\prime$ loop processes.

For both $b'$ and $t'$, there are CKM suppressed FCNC decays as
well. We consider $b^\prime\to sX$ and $t^\prime\to cX$
transitions. The unitarity relations are $\Sigma V_{is}^*V_{ib'} =
0$, and $\Sigma V_{cj}^*V_{tj} = 0$, respectively. For the loop
amplitudes, however, since to good approximation $u$ and $c$, as
well as $d$, $s$ and $b$, are practically degenerate on the $b'$,
$t'$, $t$ scale, we can take
\begin{eqnarray}
V_{us}^*V_{ub'} + V_{cs}^*V_{cb'}
 &=& -(V_{ts}^*V_{tb'} + V_{t's}^*V_{t'b'}), \\
V_{cd}V_{t'd}^* + V_{cs}V_{t's}^* + V_{cb}V_{t'b}^*
 &=& -V_{cb'}V_{t'b'}^*.
\end{eqnarray}
The loop amplitudes for $b^\prime\to s$ and $t^\prime\to c$
transitions can then be expressed to good approximation as
\begin{eqnarray}
{\cal M}_{b'\to s} & \propto &
 V_{ts}^*V_{tb'}[f(m_{t})-f(0)]+V_{t's}^*V_{t'b'}[f(m_{t'})-f(0)],
 \label{abps} \\
{\cal M}_{t'\to c} & \propto & V_{cb'}V_{t'b'}^*[f(m_{b'})-f(0)].
 \label{atpc}
\end{eqnarray}

Comparing Eq. (\ref{atpc}) with Eq. (\ref{attp}), the difference
between $t'\to c$ and $t'\to t$ is just in the CKM factor
$V_{tb'}$ and $V_{cb'}$, plus the kinematic difference of having a
light $c$ vs a heavy $t$ in the final state (thus, the functions
$f(x)$ are different).
For $b'\to s$ compared to $b'\to b$, things are more complicated,
and is more interesting. Eq.~(\ref{abbp}) should have had the same
form as Eq.~(\ref{abps}), but was simplified by the observation
that $|V_{ub}^*V_{ub'}| \ll 1$ and $|V_{cb}^*V_{cb'}| \ll 1$ are
likely, so one has $V_{tb}^*V_{tb'} \approx - V_{t'b}^*V_{t'b'}$,
and the light quark contribution can be ignored. In
Eq.~(\ref{abps}), however, one cannot ignore the light quark
effect, which provides the proper GIM subtraction for the $t$ and
$t'$ contributions. It is interesting to stress that both the CKM
coefficients $V_{ts}^*V_{tb'}$ and $V_{t's}^*V_{t'b'}$ have
nontrivial $CP$ phases, which should be in principle different.
The phase difference, together with possible absorptive parts of
the loop functions $f(m_{t})-f(0)$ and $f(m_{t'})-f(0)$, can lead
to $CP$ violation, such as $b'\to sX$ vs $\bar b'\to \bar sX$
partial rate differences.

%%%%%%%%%%%%%%%%%%%%%%%%%%%%%%
\begin{figure}[b!]
\smallskip\smallskip
%%%%%%%%%%%%%%%%%%%%%
\centerline{{ \epsfxsize3.4 in \epsffile{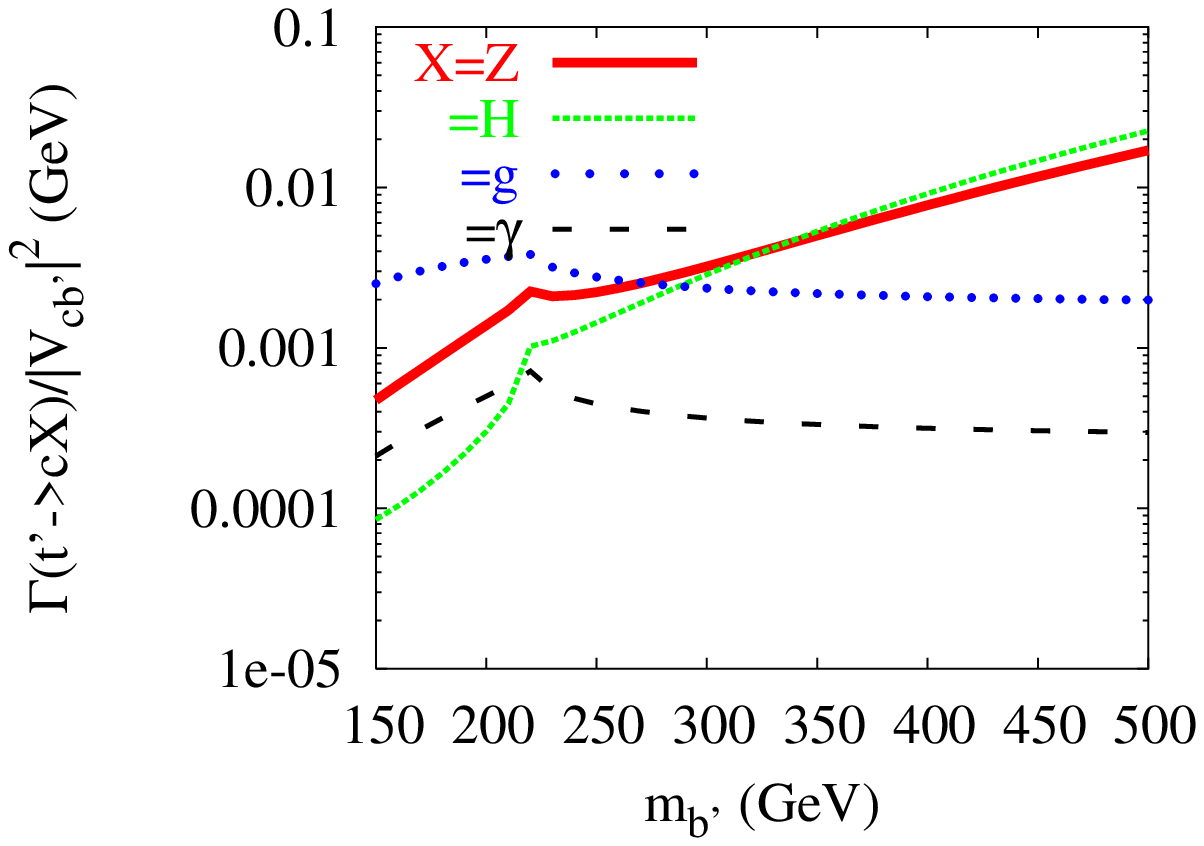}}
\hskip-1.6cm \epsfxsize3.4 in \epsffile{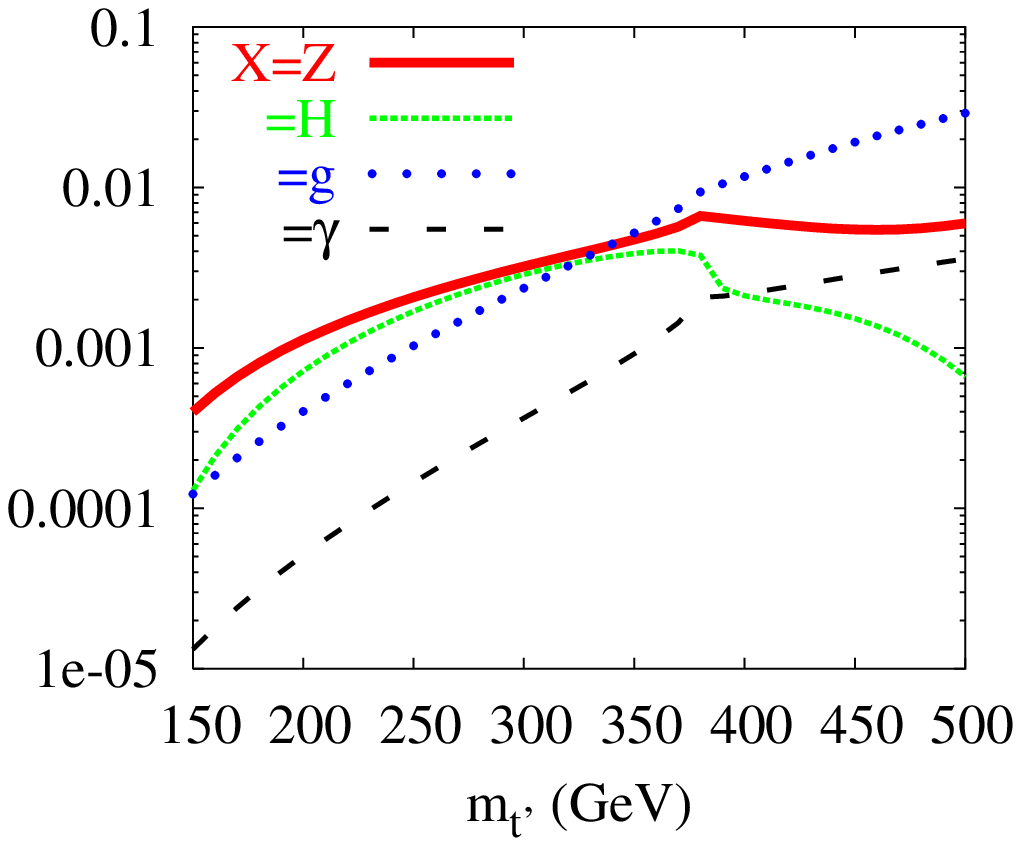} }
\smallskip\smallskip
\caption{Decay width of $\Gamma(t'\to c\{Z, H, g, \gamma\} )$
normalized to $V_{cb'}^2$ as a function of $m_{b'}$ for
$m_{t'}=300$ GeV (left), and as a function of $m_{t'}$ for
$m_{b'}=300$ GeV (right). }
\label{plot14}
\end{figure}
%%%%%%%%%%%%%%%%%%%%%%%%%%%%%%%%%%%%%%%%%

We illustrate first in Fig.~\ref{plot14} the simpler case of the
decay widths of $t'\to c$ transitions normalized to $|V_{cb'}|^2$,
as a function of $m_{b'}$ and $m_{t'}$. Comparing Fig.~\ref{plot14}
with Fig.~\ref{plot3}, it is clear that the partial widths
$\Gamma(t'\to cX)/|V_{cb'}|^2$ are slightly larger than
$\Gamma(t'\to tX)/r_{tt'}^2$. The reason is that $t'\to t$
transitions are more suppressed by phase space. As one can see
from Fig.~\ref{plot3}(b), $t'\to tZ$ and $t'\to tH$ are open only
if $m_{t'}>m_t+m_Z$ and $m_{t'}>m_t+m_H$.
It is useful to compare with the decay branching ratios given in
Fig.~\ref{plot6} for $|V_{cb'}| \simeq |V_{t's}| \simeq 0.04$. The
branching ratio for $t'\to cZ$ is about $0.00023$ times the
branching ratio of $t'\to sW$ in Fig.~\ref{plot6}(a), with $t'\to
cH$ ($cg$) just below (above), and $t'\to c\gamma$ is another
factor of 3 lower. In Fig.~\ref{plot6}(b), the branching ratios
for $t'\to cH \ga t'\to cZ$ is about $0.001$ times the branching
ratio of $t'\to sW$, with $t'\to cg$ slightly below, and $t'\to
c\gamma$ another order of magnitude lower. Depending on $r_{tt'} =
V_{tb'}V_{t'b'}^*$, the $t'\to cX$ modes could be comparable to
the $t'\to tX$ modes. For the HNS scenario \cite{HNS,HNS_KL},
where $r_{tt'}$ is not much larger than $|V_{cb'}|$, the  $t'\to
cX$ modes could dominate over $t'\to tX$ transitions. On the other
hand, if $V_{cb'} \simeq 10^{-3}$ i.e. the case of
Fig.~\ref{plot7}, then $t'\to c$ transitions should be much
suppressed compared to the $t'\to tX$ modes.

Thus, as one searches for FCNC $t'\to tX$ modes, the FCNC $t'\to
cX$ modes should not be ignored. The latter modes have the
advantage of being simpler.

For the case of $b'\to sX$, Eq.~(\ref{abps})
has both $t$ and $t'$ loop contributions, which depend on the CKM
elements $V_{ts}^*V_{tb'}$ and $V_{t's}^*V_{t'b'}$, respectively.
These elements are not the same as the elements $V_{ts}^*V_{tb}$
and $V_{t's}^*V_{t'b}$ that enter $b\to s$ transitions discussed
in Section 2. A detailed analysis would necessarily involve all
the low energy $b\to s$, $s\to d$, $b\to d$ and $c\to u$
transitions, and is clearly beyond the scope of the present paper.
Such an analysis has been performed in the HNS scenario
\cite{HNS,HNS_KL}, which is motivated by the direct $CP$ violation
problem in $B\to K^+\pi^-$ vs $K^+\pi^0$ modes, and is discussed
briefly in Sec.~6.2. It can further give \cite{HNRS} $\Delta {\cal
S} < 0$ seen in many modes at the B factories.
The $b'\to sX$ processes deserve further study, however, since
they can exhibit $CP$ violation in {\it very} heavy quark decays.
To illustrate this, we will evaluate the average partial decay
width of $b'\to sX$ defined as
\begin{eqnarray}
\overline \Gamma(b'\to s X)= \frac{1}{2} \left(\Gamma(b'\to s X) +
\Gamma(\bar{b}'\to \bar{s} X)\right),
 \label{average}
\end{eqnarray}
and the $CP$ asymmetry defined as
\begin{eqnarray}
{\cal A}_{\rm{CP}}(b'\to s X)=\frac{\Gamma(b'\to s
  X)-\Gamma(\bar{b}'\to \bar{s} X) }{\Gamma(b'\to s
  X)+\Gamma(\bar{b}'\to \bar{s} X) }.
\label{cpa}
\end{eqnarray}
Since $V_{ts}^*V_{tb'}$ and $V_{t's}^*V_{t'b'}$ would in general
have different $CP$ violating phase, while the corresponding $CP$
conserving phases in $f(m_t) - f(0)$ and $f(m_{t'}) - f(0)$ could
also be different, we expect ${\cal A}_{\rm{CP}}(b'\to s X) \neq
0$.

%%%%%%%%%%%%%%%%%%%%%%%%%%%%%%
\begin{figure}[t!]
\smallskip\smallskip
%%%%%%%%%%%%%%%%%%%%%
\centerline{{ \epsfxsize3.4 in \epsffile{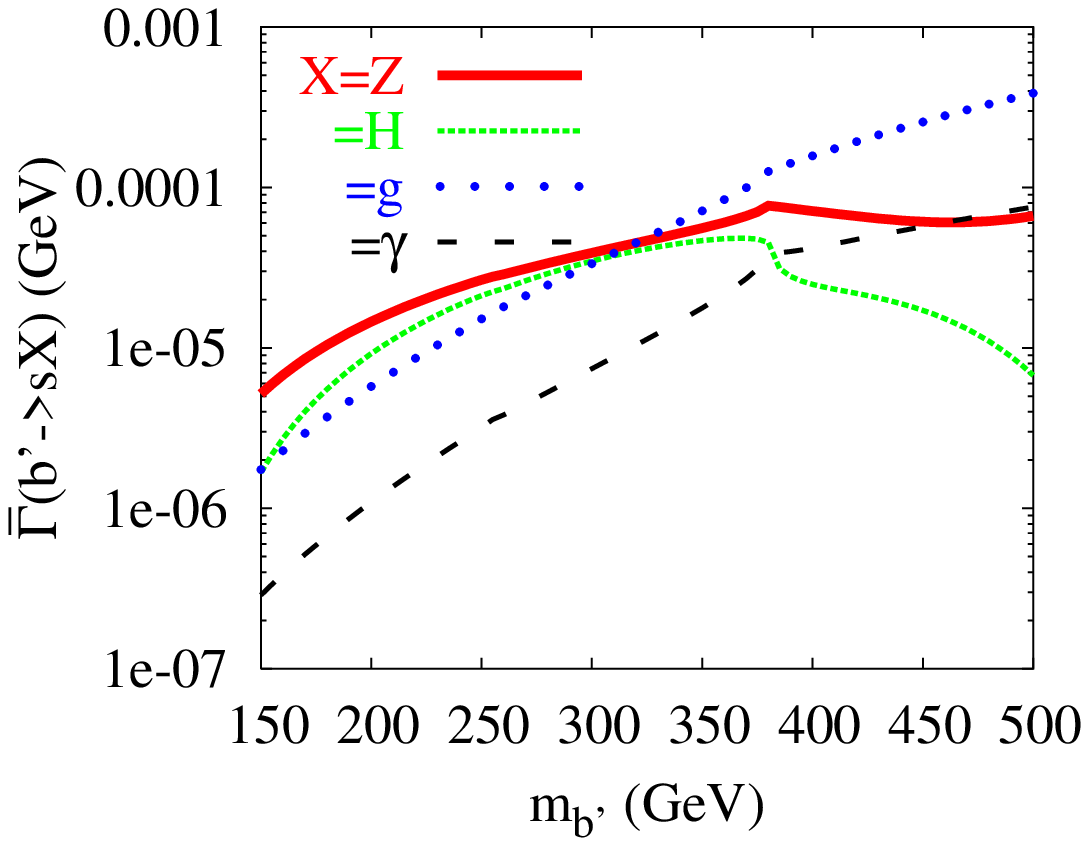}}
\hskip-1.cm \epsfxsize3.4 in \epsffile{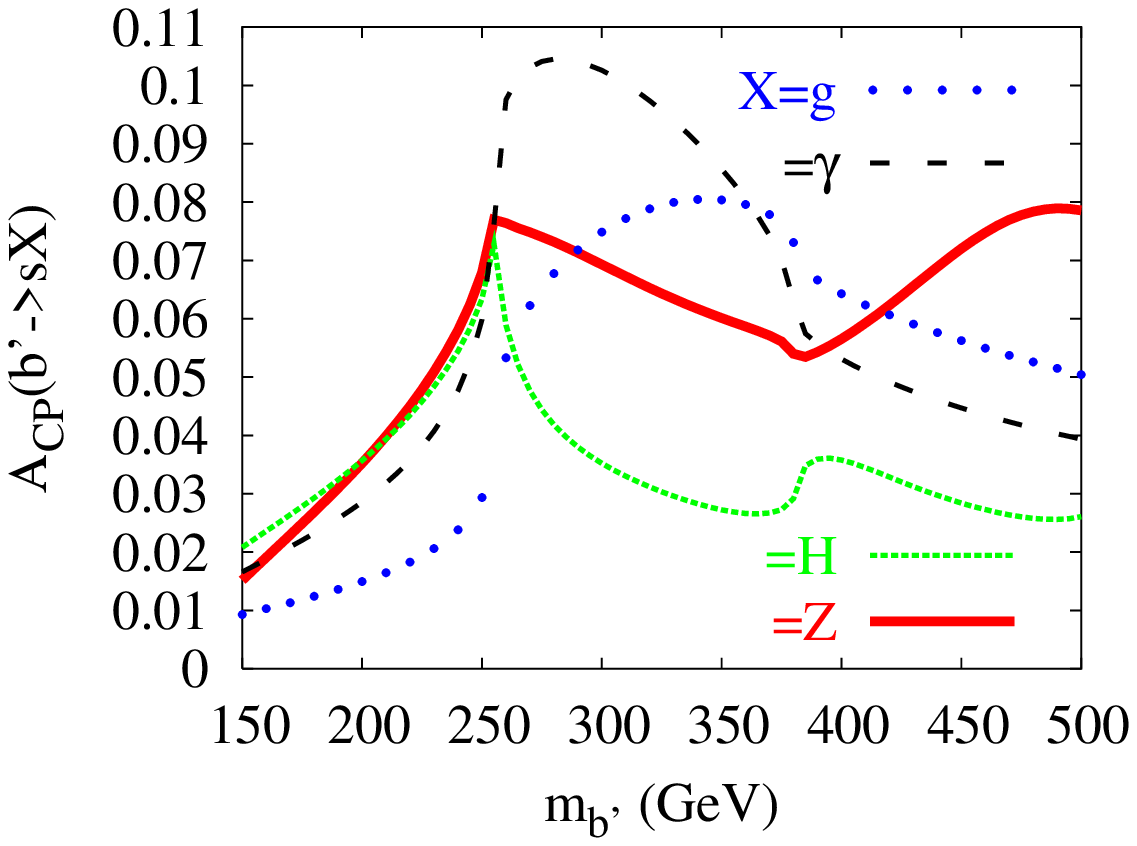} }
\smallskip\smallskip
\caption{(Left) Decay width $\overline\Gamma (b'\to s X)$ and
${\cal A}_{\rm{CP}}$ for $b'\to s X$ (right) as a function of
$m_{b'}$ for $m_{t'}=300$ GeV in the HNS scenario. }
 \label{hns}
\end{figure}
%%%%%%%%%%%%%%%%%%%%%%%%%%%%%%%%%%%%%%%%%

We use the HNS scenario as illustration, where we
find~\cite{HNS_KL}
\begin{equation}
V_{ts}^*V_{tb'} = -0.012 \, e^{i\,24^\circ}, \ \ \
V_{t's}^*V_{t'b'} = -0.11 \, e^{i\,70^\circ}.
\end{equation}
From this we expect that the size of $CP$ violation will be of the
order $|V_{ts}^*V_{tb'}/V_{t's}^*V_{t'b'}|\approx 0.11$ or less.
We illustrate in Fig.~\ref{hns} the average decay width $\overline
\Gamma(b'\to s X)$ and $CP$ asymmetry ${\cal A}_{\rm{CP}}(b'\to s
X)$ as a function of $m_{b'}$. Note that the HNS scenario fixes
$m_{t'}$ to 300 GeV, but the analysis is almost independent of
$m_{b'}$.
Comparing Fig.~\ref{hns}(a) and Fig.~\ref{plot14}(b), we see that,
up to an overall factor, they are rather similar. This is because
in the HNS scenario, the $V_{t's}^*V_{t'b'}$ term, i.e. the second
term of Eq.~(\ref{abps}), is dominant. This term is rather similar
to Eq.~(\ref{atpc}) with $m_{b'} \leftrightarrow m_{t'}$. In fact,
for Higgs and gluon final state, they should be identical. Even
for the $Z$ and $\gamma$ final state, as seen from
Fig.~\ref{hns}(a), the difference is minor.

Of course, it is the presence of the first $t$ contribution term
of Eq.~(\ref{abps}), which interferes with the second $t'$
contribution term, that makes $CP$ violation possible. As
expected, from Fig.~\ref{hns}(b) we see that the $CP$ asymmetries
for the $b'\to sX$ transitions are in the range of a few \% up to
10\%. For light $m_{b'}\approx 150-200$ GeV, ${\cal A}_{CP}$ is
rather small because $f(m_{t})-f(0)$ and $f(m_{t'})-f(0)$ have
approximately similar phase. The asymmetries rise with $m_{b'}$ as
the $b'\to tW$ threshold is  approached. Passing this threshold,
the asymmetries for the $b'\to sZ$ and $sH$ modes start to drop,
but continue to rise for the $b'\to sg$ and $s\gamma$ modes.
Crossing the $b'\to t'W$ threshold, however, the asymmetries for
the $b'\to sg$ and $s\gamma$ modes start to drop. The asymmetry
for the $b'\to sZ$ starts to rise again, but for the $b'\to sH$
mode, it drops after rising briefly.

To conclude, $b'\to sX$ rates could be comparable to $b'\to bX$ if
$V_{t's}^*V_{t'b'}$ is not much smaller than $V_{t'b}^*V_{t'b'}$,
which is the case for the HNS scenario. The HNS scenario, however,
is not optimized for $CP$ violation effect in $b'\to sX$ modes. As
the rates for these modes drop with the strength of $V_{t's}$,
much larger $CP$ violation effects could be possible. The
phenomena seem rich and deserve further study, which we refer to a
future work.

%%%%%%%%%%%%%%%%%%%%%%%%%%%%%%%%%%%%%%


\begin{thebibliography}{99}
\bibitem{vubp} $V_{t'd}$ and $V_{ub'}$ cannot both vanish, but for
purpose of $b'$ and $t'$ search, ignoring them is a valid
approximation for practical purposes.
\end{thebibliography}
\end{document}